\documentclass[10pt,twoside]{article}
\usepackage{own_sngl}
\usepackage[dvips]{graphics}
\usepackage{epsfig}
\usepackage{epsf}
\usepackage[figuresleft]{rotating}
\newcommand{\oversim}[2]{\protect{\mbox{\lower0.5ex\vbox{%
   \baselineskip=0pt\lineskip=0.2ex
   \ialign{$\mathsurround=0pt #1\hfil##\hfil$\crcr#2\crcr\sim\crcr}}}}} 
\newcommand{\simgreat}{\mbox{$\,\mathrel{\mathpalette\oversim>}\,$}} 
\newcommand{\simless} {\mbox{$\,\mathrel{\mathpalette\oversim<}\,$}} 
\slugcomment{{\em MNRAS: in press}}
\lefthead{Kroupa}
\righthead{IMF variations}
\begin{document}
%
\title {On the variation of the Initial Mass Function}
\author {Pavel Kroupa\\
\medskip
\small{Institut f\"ur Theoretische Physik und Astrophysik\\
Universit\"at Kiel, D-24098 Kiel, Germany}}
%
\begin{abstract}
\noindent 
A universal initial mass function (IMF) is not intuitive, but so far
no convincing evidence for a variable IMF exists.  The detection of
{\it systematic variations} of the IMF with star-forming conditions
would be the {\it Rosetta Stone} for star formation.

In this contribution an average or Galactic-field IMF is defined,
stressing that there is evidence for a change in the power-law index
at only two masses: near $0.5\,M_\odot$ and $0.08\,M_\odot$.  Using
this supposed universal IMF, the uncertainty inherent to any
observational estimate of the IMF is investigated, by studying the
scatter introduced by Poisson noise and the dynamical evolution of
star clusters. It is found that this apparent scatter reproduces quite
well the observed scatter in power-law index determinations, thus
defining the {\it fundamental limit} within which any true variation
becomes undetectable.  The absence of evidence for a variable IMF
means that any true variation of the IMF in well studied populations
must be smaller than this scatter.

Determinations of the power-law indices $\alpha$ are subject to
systematic errors arising mostly from unresolved binaries. The
systematic bias is quantified here, with the result that the
single-star IMFs for young star-clusters are systematically steeper by
$\Delta\alpha\approx0.5$ between $0.1$~and~$1\,M_\odot$ than the
Galactic-field IMF, which is populated by, on average, about 5~Gyr old
stars.  The MFs in globular clusters appear to be, on average,
systematically flatter than the Galactic-field IMF (Piotto \& Zoccali
1999; Paresce \& De Marchi 2000), and the recent detection of ancient
white-dwarf candidates in the Galactic halo and absence of associated
low-mass stars (M\'endez \& Minniti 2000; Ibata et al. 2000) suggests
a radically different IMF for this ancient population. Star-formation
in higher-metallicity environments thus appears to produce relatively
more low-mass stars.  While still tentative, this is an interesting
trend, being consistent with a systematic variation of the IMF as
expected from theoretical arguments.
\end{abstract}

\vskip 5mm
%

\section{INTRODUCTION} 
\label{sec:intro}
\noindent
Fundamental arguments suggest that the initial mass function (IMF)
{\it should} vary with the pressure and temperature of the
star-forming cloud in such a way that higher-temperature regions ought
to produce higher average stellar masses (Larson 1998). This is
particularly relevant to the formation of population~III stars,
because the absence of metals and more intense ambient radiation field
imply higher temperatures.

The IMF inferred from Galactic-field star-counts can be conveniently
described by a 3--4 part power-law (eqs.~{\ref{eq:xi} and~\ref{eq:imf}
below). The Galactic field was populated by many different
star-formation events. Given this well-mixed nature of the solar
neighbourhood, present-day star-formation ought to lead to {\it
variations about the Galactic-field IMF}. In particular, a systematic
difference ought to be evident between low-density environments
(e.g. Taurus--Auriga; $\rho$~Oph) and high-density regions (e.g. Orion
Nebula Cluster, ONC), because above a certain critical density,
star-forming clumps interact with each other before their collapse
completes (Murray \& Lin 1996; Allen \& Bastien 1995; Price \&
Podsiadlowski 1995; Klessen \& Burkert 2000). On considering the ratio
between the fragment collapse time and the collision time-scale and
applying the analysis of Bastien (1981, his eq.~8), it becomes
apparent that the IMF in clusters similar to $\rho$~Oph cannot be
shaped predominantly through collisions between collapsing clumps.
This is supported through the finding by Motte, Andr\'e \& Neri (1998)
that the pre-stellar-clump MF in $\rho$~Oph is similar to the observed
MF for pre-main sequence stars in $\rho$~Oph. It is somewhat steeper
than the Galactic-field IMF, especially if the binary systems that
must be forming in the pre-stellar cores are taken into account.
Noteworthy is that both, the pre-stellar clump MF and the
Galactic-field IMF have a reduction of the power-law index below about
$0.5\,M_\odot$.  In the core of the ONC, however, pre-stellar cores
most likely did interact significantly (Bonnell, Bate \& Zinnecker
1998; Klessen 2001).  Furthermore, once the OB stars ignite in a
cluster such as the ONC, they have a seriously destructive effect
through the UV flux, strong winds and powerful outflows, and so are
likely to affect the formation of the least massive objects, including
planets. This can happen for example through destruction of the
accretion envelope, so that extreme environments like the Trapezium
Cluster may form a surplus of unfinished stars (brown dwarfs, BDs)
over Taurus--Auriga. Luhman (2000) finds empirical evidence for this,
but detailed dynamical modelling is required to exclude the
possibility raised here that at least part of this difference may be
due to the disruption of BD--BD and star--BD binaries in a dynamically
evolved population such as the Trapezium Cluster.

A conclusive difference has not been found between the IMF in
Taurus--Auriga (Kenyon \& Hartmann 1995; Briceno et al. 1998) and
$\rho$~Oph (Luhman \& Rieke 1999) on the one hand, and the ONC (Palla
\& Stahler 1999; Muench, Lada \& Lada 2000; Hillenbrand \& Carpenter
2000) on the other. Similarly, Luhman \& Rieke (1998) point out that
no significant IMF differences for pre-main sequence populations
spanning two orders of magnitude in density have been found.  Such
conclusions rely on pre-main sequence tracks that are unreliable for
ages younger than about 1~Myr (I.~Baraffe, priv.commun.), because the
density, temperature and angular momentum distribution within the
pre-main sequence star is likely to remember the accretion history
(Wuchterl \& Tscharnuter 2000).  Nevertheless, in support of the
universal-IMF notion, it is remarkable how similar the Galactic field
MF is to the MF inferred for the Galactic bulge (Holtzman et al. 1998;
Zoccali et al. 2000), again with a flattening around
$0.5\,M_\odot$. Presumably star-formation conditions during bulge
formation were quite different to the conditions witnessed in the
Galactic disk, but the bulge and disk metallicities are
similar. Further related discussions on this topic can be found in
Gilmore \& Howell (1998).

The quest for detecting variations in the IMF has been significantly
pushed forward by Scalo (1998), who compiled determinations of the
logarithmic power-law index, $\Gamma$ (eq.~\ref{eq:xiL}), for many
clusters and OB associations in the Milky Way (MW) and the Large
Magellanic Cloud (LMC), which has about $1/5$th$-1/3$rd the
metallicity of the MW (e.g. Holtzman et al. 1997).  While no
systematic variation is detectable in a plot of $\Gamma$ against
stellar mass, $m$, between populations belonging to the two galaxies,
a large constant scatter in $\Gamma$ for stars more massive than
$1\,M_\odot$ is evident instead. This raises the question of how large
{\it apparent IMF variations} are due to small number statistics and
other as yet unexplored observational uncertainties, and if this noise
can mask, or even render undetectable, any true variations of the IMF.

Elmegreen (1999) shows that statistical variations of $\Gamma$, that
are not dissimilar to the observed ones, result naturally from a model
in which the Salpeter IMF constructs from random sampling of
hierarchically structured clouds, if about $N=100$ stars are observed.
This model predicts that the scatter in $\Gamma$ must decrease with
increasing $N$.

In this contribution the reductionist philosophy is followed according
to which all non-star-formation sources of {\it apparent variations}
of the IMF must be understood before the spread of $\Gamma$ can be
interpreted as being due to the star formation process.  To achieve
this, an {\it invariant IMF is assumed} to study three possible
contributions to the large scatter seen in the alpha-plot: {\bf (i)}
Poisson scatter due to the finite number of stars in a sample. This is
similar to Elmegreen's approach, except that no explicit link to the
distribution of gas clumps is made.  {\bf (ii)} Loss of stars of a
preferred mass-scale as their parent star-clusters evolve
dynamically. This dynamical loss is not a simple function of stellar
mass, because of the complex stellar-dynamical events in a young
cluster. For example, while low-mass members preferably diffuse
outwards as a result of energy equipartition, massive stars sink
inwards where they meet and expel each other rather
effectively. Finally, {\bf (iiia)} wrong mass estimates because most
stars are born in binary systems, and observers usually cannot resolve
the systems.  The simplest approach, taken here, is to replace the two
component masses by the combined mass of the binary system, and to
measure the system MF.

Issues also contributing to the scatter but not dealt with here are
the following: {\bf (iiib)} An observer infers the mass of a star from
the observed luminosity incorrectly if the star is an unresolved
binary, {\bf (iiic)} wrong mass estimates from luminosities in the
event of higher-order multiplicities, which is a major bias for
massive stars (e.g. Preibisch et al. 1999), and {\bf (iva)} stellar
evolution and the application of incorrect pre-main sequence and
main-sequence evolutionary tracks, which corrupts the masses inferred
from observed quantities as the stars evolve to or along the main
sequence, and {\bf (ivb)} incorrect estimates of stellar masses as a
result of rapidly rotating massive stars and the use of non-rotating
stellar evolution models.  One issue to be stressed in this connection
is that {\it stellar evolution} theory retains significant
uncertainties (Kuruzc 2000; Maeder \& Meynet 2000; Heger \& Langer
2000), which can only be reduced through continued attention.

The present study thus probably {\it underestimates} the scatter by
focusing on points~(i) to~(iiia), but allows an assessment of the {\it
fundamental limits} within which apparent IMF variations swamp true
variations.

The {\it alpha-plot} and the form of the universal IMF adopted here
are introduced in Section~\ref{sec:imf}, and statistical variations of
the power-law index are studied in Section~\ref{sec:sv}. The
star-cluster models are described in Section~\ref{sec:clmods}, and
Section~\ref{sec:res} contains the results on the variation of the
MF. In Section~\ref{sec:disc} the {\it dichotomy} in the alpha-plot
and available evidence for a truly variable IMF are discussed. The
conclusions are presented in Section~\ref{sec:concl}.

\section{THE ALPHA-PLOT AND THE GALACTIC-FIELD IMF}
\label{sec:imf}
\noindent 
Observational data in the alpha-plot are used to infer a universal IMF.

\subsection{The alpha--plot}
\label{sec:apl}
Scalo (1998) combined available IMF estimates for star clusters and
associations by plotting the power-law index, $\Gamma$
(eq.~\ref{eq:xiL} below), against the mean log$_{10}m$ of the mass
range over which the index is measured (his fig.~5).
Fig.~\ref{fig:a_m} shows these {\it same data} by plotting the
power-law index $\alpha = 1-\Gamma$ against log$_{10}m_{\rm av}$. The
{\it alpha-plot} clearly shows the flattening of the IMF for
$m\simless 0.5\,M_\odot$. It also shows no systematic difference
between MW and LMC populations, as already shown by Massey et
al. (1995b) for massive stars.  This is also verified for $0.6\simless
m\simless 1.1\,M_\odot$ by Holtzman et al. (1997), who use deep HST
photometry for LMC fields and apply Monte Carlo models that include
binary systems, various star-formation histories (sfh) and
metallicities as well as observational errors.

\begin{figure}
\begin{center}
\rotatebox{-90}{\resizebox{0.77 \textwidth}{!}
{\includegraphics{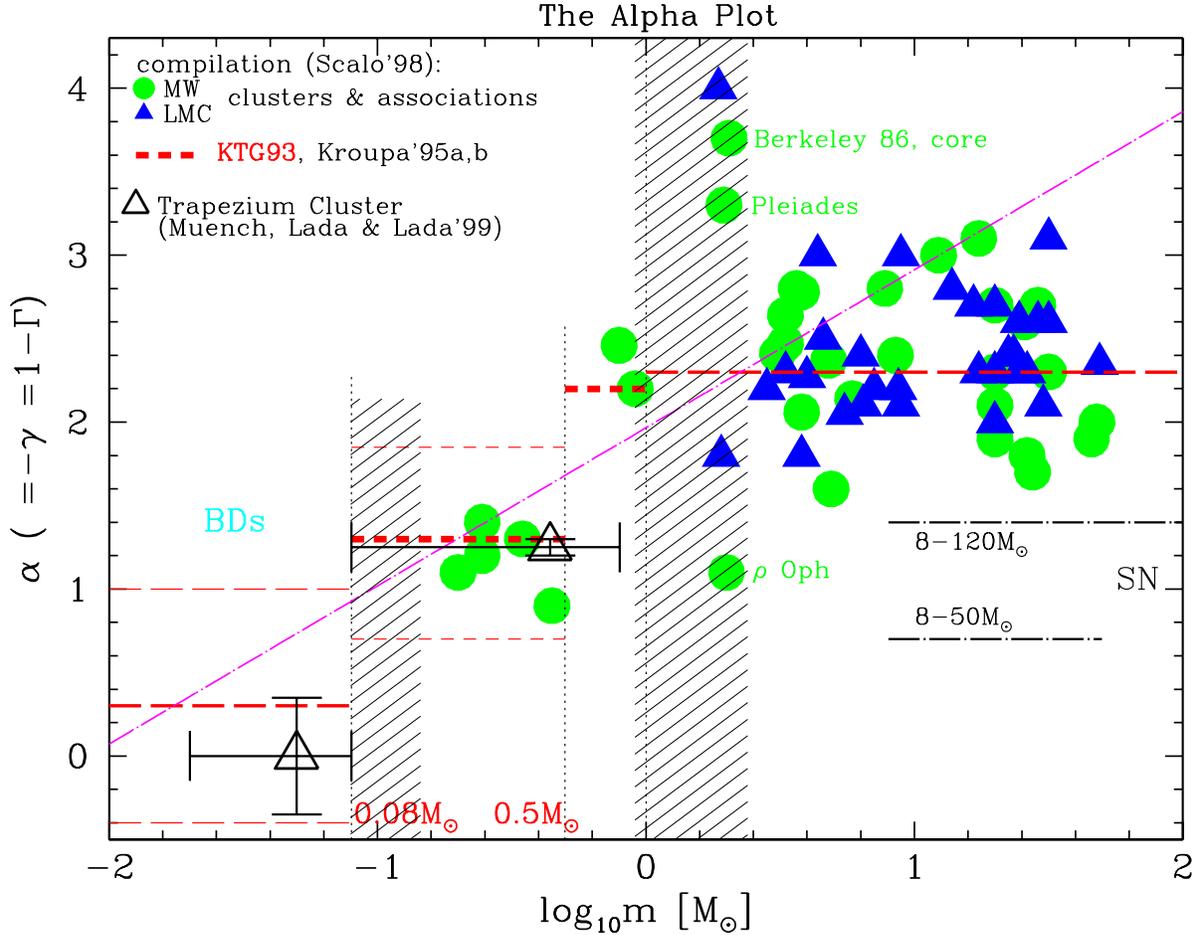}}}
\vskip 0mm
\caption
{\small{ The {\it alpha-plot}.  The data show the compilation by Scalo
(1998) of determinations of $\alpha$ over different mass ranges for
Milky-Way (MW) and Large-Magellanic-Cloud (LMC) clusters and OB
associations.  Unresolved multiple systems are not corrected for.  The
large open triangles (Muench, Lada \& Lada 2000 from Orion Nebula
Cluster observations, binary corrections not applied) serve to
illustrate the present knowledge for $m<0.1\,M_\odot$.  The horizontal
long-dashed lines in the BD regime are the Galactic-field IMF
(eq.~\ref{eq:imf}) with associated approximate uncertainties. For
$0.08\le m \le 1.0\,M_\odot$ the thick short-dashed lines represent
the KTG93 single-star IMF (Kroupa, Tout \& Gilmore 1993), which has
$\alpha_3=2.7$ for $m>1\,M_\odot$ from Scalo's (1986)
determination. The long-dashed lines for $m>1\,M_\odot$ show the
approximate average $\alpha=2.3$, which is adopted in the
Galactic-field IMF (eq.~\ref{eq:imf}).  The Miller \& Scalo (1979)
log-normal IMF for a constant star-formation rate and a Galactic disk
age of 12~Gyr is plotted as the diagonal long-dash-dotted line.  The
long-dash-dotted horizontal lines labelled ``SN'' are those
$\alpha_3=0.70 (1.4)$ for which 50~\% of the stellar (including BD)
mass is in stars with $8 - 50 (8 - 120)\,M_\odot$.  The vertical
dotted lines delineate the four mass ranges (eq.~\ref{eq:imf}), and
the shaded areas highlight those stellar mass regions where the
derivation of the IMF is additionally complicated due to unknown ages,
especially for Galactic field stars: for $0.08<m< 0.15\,M_\odot$
long-contraction times make the conversion from an empirical LF to an
IMF dependent on the precise knowledge of the age, and for $0.8<
m<2.5\,M_\odot$ post-main sequence evolution makes derived masses
uncertain in the absence of precise age knowledge. A few of the MW
data are labelled by their star-clusters, and Table~\ref{tab:a_m}
lists the $m_{\rm av}<1\,M_\odot$ data.  }}
\label{fig:a_m}
\end{center}
\end{figure}

The models discussed in Section~\ref{sec:res} show that unresolved
binary systems mostly affect the region $m\simless 1\,M_\odot$, the
data for which are listed in Table~\ref{tab:a_m}. Perusal of the
references shows that only Meusinger et al. (1996) attempted a
correction for unresolved binary systems. However, they adopted an
artificial model of Reid (1991; see discussion in Kroupa 1995a), in
which the binary proportion is only 40~per~cent, half of which have
similar companion masses. This is an unrealistic model (K\"ahler
1999), and leads to essentially no difference between the system and
single-star LFs (fig.~7 in Meusinger et al. 1996, compare with fig.~11
in Kroupa 1995d).  Their binary correction can thus be safely
neglected.

\begin{table}
{\small
\begin{minipage}[t]{10cm}
\vskip 5mm
\begin{tabular}{ccll}

log$_{10}m_{\rm av}$ &$\alpha$ &cluster &ref.\\

$-0.70$  &1.10 &$\rho$~Oph &(Williams et al. 1995b)\\

$-0.61$  &1.40 &$\rho$~Oph &(Comeron, Rieke \& Rieke 1996)\\

$-0.61$  &1.20 &NGC~2024 &(Comeron, Rieke \& Rieke 1996)\\

$-0.46$  &1.30 &Praesepe &(Williams, Rieke \& Stauffer 1995a)\\

$-0.35$  &1.10  &Pleiades &(Meusinger et al. 1996)\\

$-0.10$  &2.46 &ONC &(Hillenbrand 1997)\\

$-0.04$  &2.20 &Praesepe &(Williams, Rieke \& Stauffer 1995a)\\

\end{tabular}
\end{minipage}
}
\caption{\small{The data from Scalo's (1998) compilation with $m_{\rm
av}<1\,M_\odot$. }}
\label{tab:a_m}
\end{table}

\subsubsection{ $m>3\,M_\odot$ }
\label{sec:ms}
For $m\simgreat 3\,M_\odot$ the data suggest that the Salpeter
power-law value, $\alpha=2.3$, is a reasonable fit over the whole
range, as is also stressed by Massey (1998). Massey \& Hunter (1998),
for example, deduce that $\alpha\approx2.3$ for $2.8<m<120\,M_\odot$
in the massive cluster R136 in the LMC.  This value is thus adopted
throughout the rest of this paper, although notable examples of exotic
clusters exist. The two massive, apparently young (2--4~Myr) Arches
and Quintuplet clusters lying very close to the Galactic centre
(projected distance 30~pc) have $\alpha\approx1.65$ (Figer et
al. 1999), and the Galactic star-burst template cluster NGC~3603 is
found to have $\alpha\approx1.7$ (Eisenhauer et al. 1998). Further
work is desired to establish the exact nature of the central clusters,
and clarify the age discrepancy between the low-mass and massive stars
noted for NGC~3603, a problem possibly associated with pre-main
sequence models.

It is important to keep in mind that $\alpha$ may be systematically
steeper than $\alpha=2.3$ (or 1.7) due to unresolved binary systems,
which are not usually corrected for in IMF estimates.  The
multiplicity proportion of massive stars is very high (Mason et
al. 1998). For example, Preibisch et al. (1999) find that the OB stars
in the well studied ONC have, on average, 1.5 companions.  For each
primary, there is thus usually more than one secondary that adds at
lower masses, steepening the observed IMF when corrected for.  The
effect depends on $\alpha$, and Sagar \& Richtler (1991) calculate
that $\Delta\alpha=+0.34$ for $\alpha=2.5$ and a binary proportion
$f=0.5$ (eq.~\ref{eq:f} below). If $f=1$ (each massive primary has 1.0
companions) they obtain $\Delta\alpha=+0.40$. $\Delta\alpha$ is likely
to be larger still, because each massive primary probably has more
than one companion, typically. Since the empirical data in
Fig.~\ref{fig:a_m} implies an average $\alpha\approx2.3$ for
$m\simgreat 3\,M_\odot$, the true single-star IMF may in fact have
$\alpha\approx2.7 (=2.3+0.4)$, or even larger. A similar conclusion is
reached by Scalo (1998, at the end of his section~4). Such corrections
will not be removed if spectroscopic mass determinations are used
instead of the inferior mass-estimates using photometry (Massey et
al. 1995a), since unresolved systems will have similar effects on a
spectroscopic sample.

In this paper the approximate average $\alpha=2.3$ is adopted, with
the aim of studying the effect of unresolved binary systems on the
$\alpha$ inferred from the {\it system MF}, which an observer would
deduce from the mixture of single stars and binary systems in a
population resulting from star-cluster evolution with initially
$f=1$. Because the assumptions (Section~\ref{sec:pop}) imply that
massive stars have very-low mass companions in this model, and because
only binary systems are searched for in the data reduction software,
the resulting model bias will be an underestimate. Further work is
necessary to address this particular issue, which is also discussed
further at the end of Section~\ref{sec:sv} and in
Section~\ref{sec:nb}.

The remarkable feature for $m\simgreat 3\,M_\odot$ in the alpha plot
is the {\it constant scatter}, and that the various power-law indices
are distributed more or less randomly throughout the region
$\alpha=2.3\pm0.7$ without a significant concentration towards some
value.

\subsubsection{ $0.8< m< 3\,M_\odot$ }
The region for $0.8\simless m\simless 3\,M_\odot$ shows an unusually
large scatter. It is shaded because this particular mass range is
problematical for a number of reasons. 

Analysis of Galactic-field star-counts run into the difficulty that
the age of the Galactic disk is comparable to the life-time of these
stars so that stellar evolution corrections become very significant,
but for this the sfh must be known (Scalo 1986; Haywood, Robin \&
Cr\'ez\'e 1997).  That interesting constraints can be placed on the MW
IMF using independently derived sfhs is shown by Maciel \& Rocha-Pinto
(1998), where the problems associated with the estimation of the
field-IMF for massive stars are documented.  

The large spread of the cluster values in the region $0.8\simless
m\simless 3\,M_\odot$ may be due to the fact that the observed
clusters have ages such that the stars in this mass range count to the
most massive remaining in the clusters. They are thus subject to
advanced stellar evolution and/or dynamical ejection from the cluster,
because the most massive stars usually interact in the vicinity of the
cluster core.  Which of these is applicable is a sensitive function of
the age of the cluster and the number of stars in it (more on this in
Section~\ref{sec:amclpop}). Finally, stellar evolution is by no means
a solved subject for stars in this mass range (Dominguez et al. 1999)
with remaining significant uncertainties. This compromises the
conversion of stellar luminosity to mass. Ignoring the large scatter
in this mass range, it can be seen that a single power-law index
becomes applicable for $m>0.5\,M_\odot$.

\subsubsection{ $0.1< m< 1\,M_\odot$ }
The Galactic-field single-star IMF fits the data shown in
Fig.~\ref{fig:a_m} exceedingly well for $0.1<m<1\,M_\odot$ (that this
agreement may be fortuitous though is shown in
Section~\ref{sec:revimf}). In particular, it is remarkable that the
data suggest a change in $\alpha$ near $0.5\,M_\odot$, as was
initially derived from solar-neighbourhood star-counts by Kroupa, Tout
\& Gilmore (1991, hereinafter KTG91), and later confirmed by Kroupa,
Tout \& Gilmore (1993, hereinafter KTG93) and Kroupa (1995a) using a
different mass--luminosity relation, a much more detailed star-count
analysis including main-sequence and pre-main sequence stellar
evolution, and with different statistical tests. Similar work by
Gould, Bahcall \& Flynn (1997) using HST star-counts and Reid \& Gizis
(1997), who study a proposed extension of the nearby stellar sample to
somewhat larger distances, also confirm these findings, as do Chabrier
\& Baraffe (2000), who estimate $\alpha\approx1.2\pm0.1$ using the
nearby volume-limited LF.

Of special importance is the mass range $0.5-1\,M_\odot$: The local
sample of known stars is sufficiently large in this mass range that
the nearby volume-limited LF is very well defined (Kroupa
2001a). Also, unresolved binaries do not significantly affect the LF
in this mass range, because the stellar sample does not contain stars
with $m>1\,M_\odot$ that can hide lower-mass companions.  The
mass--luminosity relation is also well understood for these stars, so
that the MF determination should be accurate and precise. It is not
surprising that the power-law slope has changed little over the
decades (Salpeter 1955: $\alpha=2.35$ for $0.4<m<10\,M_\odot$). From
Fig.~\ref{fig:a_m} an uncertainty of $\alpha=2.3\pm0.3$ is adopted.

Unfortunately, the local sample of stars with $m\simless0.5\,M_\odot$
is incomplete for distances larger than $d\approx5$~pc, in
contradiction to the belief by Reid \& Gizis (1997), who use
spectroscopic parallax measurements to extend their proposed
volume-limited sample using previously-known stars. Malmquist bias
pollutes their sample by multiple systems that are much further away.
The seriousness of the incompleteness of the nearby stellar census is
shown by Henry et al. (1997), and is also pointed out by Chabrier \&
Baraffe (2000).  This situation can {\it only} be improved with
large-scale and deep surveys that find candidate nearby M~dwarfs with
subsequent {\it trigonometric} parallax measurements to affirm the
distance, such that a {\it volume limited sample} can be constructed.
This will be possible through the upcoming astrometric space missions
DIVA (R\"oser 1999) and GAIA (Gilmore et al. 1998).  Being aware of
this situation, the KTG studies combined the local ($d\le5.2$~pc) {\it
volume limited} sample with {\it flux limited} deep photometric
surveys, performing detailed Monte Carlo modelling of both
Galactic-field samples. This {\it pedantic separation of the two
star-count samples} is necessary as completely different biases and
errors operate. The result is the conservative uncertainty range of
$\alpha=1.3\pm0.5$ for $0.08-0.5\,M_\odot$ (KTG93).  That the
Galactic-bulge MF shows an indistinguishable behaviour to the
Galactic-field MF in this mass range was already pointed out in
Section~\ref{sec:intro}.

\subsubsection{ $m<0.08\,M_\odot$ }
For sub-stellar masses the constraints have improved dramatically in
the past few years as a result of the significant observational effort
and instrumental advances. In the ONC, Muench, Lada \& Lada (2000) and
Hillenbrand \& Carpenter (2000) find $-1\simless\alpha\simless 1$,
although the pre-main sequence tracks are unreliable at these ages.
Similarly, in $\rho$~Oph Luhman \& Rieke (1999) estimate
$\alpha\approx0.5$, which is also found for IC348 by Najita, Tiede \&
Carr (2000).  In the Pleiades Cluster, Martin et al. (2000) estimate
$\alpha\approx0.53$, and for the solar neighbourhood, Reid et
al. (1999) quote $1\simless\alpha\simless2$, whereas Herbst et
al. (1999) estimate $\alpha<0.8$ with 90~per cent confidence on the
basis of no detections but accounting correctly for Galactic
structure. For the time-being, $\alpha=0.3\pm0.7$ is a reasonable
description of the IMF for BDs, and it will be shown in
Section~\ref{sec:amclpop} that the observed MF depends sensitively on
the dynamical age of the population. 

The region $0.08-0.15\,M_\odot$ is shaded in Fig.~\ref{fig:a_m} to
emphasise the uncertainties plaguing Galactic-field star-count
interpretations as a result of the long pre-main sequence contraction
times for these stars. As with the $0.8-3\,M_\odot$ region, the sfh
must be known. The sfh has most recently been constrained by
Rocha-Pinto et al (2000).

\subsection{The universal IMF}
\label{sec:unimf}
The available constraints can be conveniently summarised by the
multiple-part power-law IMF (see Kroupa 2001b for details),
\begin{equation}
\xi(m) \propto m^{-\alpha_i} = m^{\gamma_i},
\label{eq:xi}
\end{equation}
where
\begin{equation}
          \begin{array}{l@{\quad\quad,\quad}l}
\alpha_0 = +0.3\pm0.7   &0.01 \le m/M_\odot < 0.08, \\
\alpha_1 = +1.3\pm0.5   &0.08 \le m/M_\odot < 0.50, \\
\alpha_2 = +2.3\pm0.3   &0.50 \le m/M_\odot < 1.00, \\
\alpha_3 = +2.3\pm0.7   &1.00 \le m/M_\odot,\\
          \end{array}
\label{eq:imf}
\end{equation}
and $\xi(m)\,dm$ is the number of {\it single stars} in the mass
interval $m$ to $m+dm$. The uncertainties correspond approximately to
99~per~cent confidence intervals for $m\simgreat0.5\,M_\odot$
(Fig.~\ref{fig:a_m}), and to a 95~per cent confidence interval for
$0.1-0.5\,M_\odot$ (KTG93). Below $0.08\,M_\odot$ the confidence range
is not well determined.

Note that this form differs from Scalo's (1998) recommendation, mostly
because the correct structure in the luminosity function below
$1\,M_\odot$ is accounted for here. There is evidence for {\it only
two changes in the power-law index}, namely near $0.5\,M_\odot$ and
near $0.08\,M_\odot$.  The frequently used Miller \& Scalo (1979) IMF
fails in the region $0.5-1\,M_\odot$, and especially for
$m\simgreat5\,M_\odot$ (Fig.~\ref{fig:a_m}, see also
Fig.~\ref{fig:imfv} below).  A useful representation of the IMF is
achieved via the {\it logarithmic} form,
\begin{equation}
\xi_{\rm L}(m)={\xi(m)\,{\rm ln}10\,m},
\label{eq:xiL}
\end{equation}
where $\xi_{\rm L}\,d{\rm log}_{10}m\propto m^{\Gamma_i}\,d{\rm
log}_{10}m = m^{-x_i}\,d{\rm log}_{10}m$ is the number of stars in the
logarithmic mass interval log$_{10}m$ to log$_{10}m+d{\rm log}_{10}m$.

The adopted IMF (eq.~\ref{eq:imf}) has a mean stellar mass
$<m>=0.36\,M_\odot$ for stars with $0.01\le m\le50\,M_\odot$, and
leads to the following stellar population: 37~\% BDs
($0.01-0.08\,M_\odot$) contributing 4.3~\% to the stellar mass, 48~\%
M~dwarfs ($0.08-0.5\,M_\odot$) contributing 28~\% mass, 8.9~\%
``K''~dwarfs ($0.5-1.0\,M_\odot$) contributing 17~\% mass, 5.7~\%
``intermediate mass (IM) stars'' ($1.0-8.0\,M_\odot$) contributing
34~\% mass, and 0.37~\% ``O'' stars ($>8\,M_\odot$) contributing 17~\%
mass. 

A remarkable property of eq.~\ref{eq:imf} is that 50~per~cent of the
mass is in stars with $0.01\le m\le1\,M_\odot$ and 50~per~cent in
stars with $1-50\,M_\odot$. Also, if $\alpha_4=0.70$ ($m>8\,M_\odot$)
then 50~per~cent of the mass is in stars with $8\le m\le50\,M_\odot$,
whereas $\alpha_4=1.4$ implies 50~per~cent mass in $8-120\,M_\odot$
stars. These numbers are useful for the evolution of star clusters,
because supernovae (SN) lead to rapid mass loss which can unbind a
cluster if too much mass resides in the SN precursors.  This is the
case in clusters that have $\alpha_3=1.80$: stars with
$8<m\le120\,M_\odot$ contain 53~per cent of the mass in the stellar
population! It is interesting that $\alpha\approx1.8$ for
$m\simgreat1\,M_\odot$ forms the lower bound on the empirical data in
Fig.~\ref{fig:a_m}. But even 'normal' ($\alpha=2.3$) star clusters
suffer seriously through the evolution of their $m>1\,M_\odot$ stars
(de La Fuente Marcos 1997).

\section{PROCEDURE AND STATISTICAL VARIATION}
\label{sec:sv}
\noindent 
One contribution to the scatter seen in the alpha-plot
(Fig.~\ref{fig:a_m}) is Poisson noise. This can be studied by sampling
$N$ stars from the adopted IMF (eq.~\ref{eq:imf}), and studying the
variation of $\alpha$ with $N$.

In order to construct synthetic alpha-plots, the following procedure
is adopted. $N$ masses are obtained by randomly sampling
eq.~\ref{eq:imf} with lower mass limit $m_{\rm l}=0.01\,M_\odot$ and
upper mass limit $m_{\rm u}=50\,M_\odot$. This upper mass limit is
chosen for consistency with the stellar-dynamical models
(Section~\ref{sec:clmods}).  The MF is constructed by binning the
masses, $m$, into~30 log$_{10}m$ bins which subdivide the range
$-2.1\le {\rm log}_{10}m \le +2.1$. Power-laws are fitted using
weighted linear regression (e.g. Press et al. 1994) to sub-ranges that
are defined as follows:
\begin{equation}
          \begin{array}{ll@{\quad\quad,\quad}l}
b1 & 6 & 
{\rm log}_{10}(0.01) < {\rm l}m \le {\rm log}_{10}(0.08),\\
b2 & 6 &
{\rm log}_{10}(0.08) < {\rm l}m \le {\rm log}_{10}(0.50),\\
b3 & 4 &
{\rm log}_{10}(0.40) < {\rm l}m \le {\rm log}_{10}(1.20),\\
b4 & 4 &
{\rm log}_{10}(1.00) < {\rm l}m \le {\rm log}_{10}(3.50),\\
b5 & 4 &
{\rm log}_{10}(3.00) < {\rm l}m \le {\rm log}_{10}(9.00),\\
b6 & 8 &
{\rm log}_{10}(5.00) < {\rm l}m,\\
          \end{array}
\label{eq:bin}
\end{equation}
where ${\rm l}m={\rm log}_{10}m/M_\odot$, and the numbers, $nb$,
behind the mass-range number (e.g. $b1$) are the number of mass bins
in the histogram in that particular mass range (e.g. $nb_1=6$).  This
sub-division ensures that the different mass regions in which
$\alpha_i$ is known to be constant (eq.~\ref{eq:imf}) are not mixed
up, but also allows studying the fitted $\alpha$ at values of ${\rm
l}m$ where, for example, stellar evolution and/or dynamical effects
are expected to be important. The result is $\alpha({\rm l}m_{\rm
av})$, where ${\rm l}m_{\rm av}$ is the average of ${\rm l}m$ over the
$nb_j$ ($j=1,6$) bins. In cases where the number of stars is too
small, or the highest mass star is less massive than
$10^{2.1}\,M_\odot$, some of the highest-mass bins remain empty,
causing ${\rm l}m_{\rm av}$ in mass-range $b6$ to vary between
renditions.

The IMF is plotted together with two renditions using $N=10^3$ stars
in Fig.~\ref{fig:mf_sv}, to illustrate the procedure.  The resulting
alpha-plot is shown in Fig.~\ref{fig:a_m_sv} for many more renditions
and different $N$.  The input IMF is obtained essentially exactly for
$N=10^6$ and~$10^5$, verifying the procedure.  The figure shows that
deviations begin to occur for $N=10^4$ in the two highest mass-ranges
($b5$ and~$b6$), because these contain only a few per~cent of $N$,
i.e. a few hundred stars, spread over about 10~mass bins. For smaller
$N$ the scatter of $\alpha({\rm l}m_{\rm av})$ becomes larger, with
the average reproducing the IMF except when the MF is under-sampled at
large masses.
\begin{figure}
\begin{center}
\rotatebox{-90}{\resizebox{0.6 \textwidth}{!}
{\includegraphics{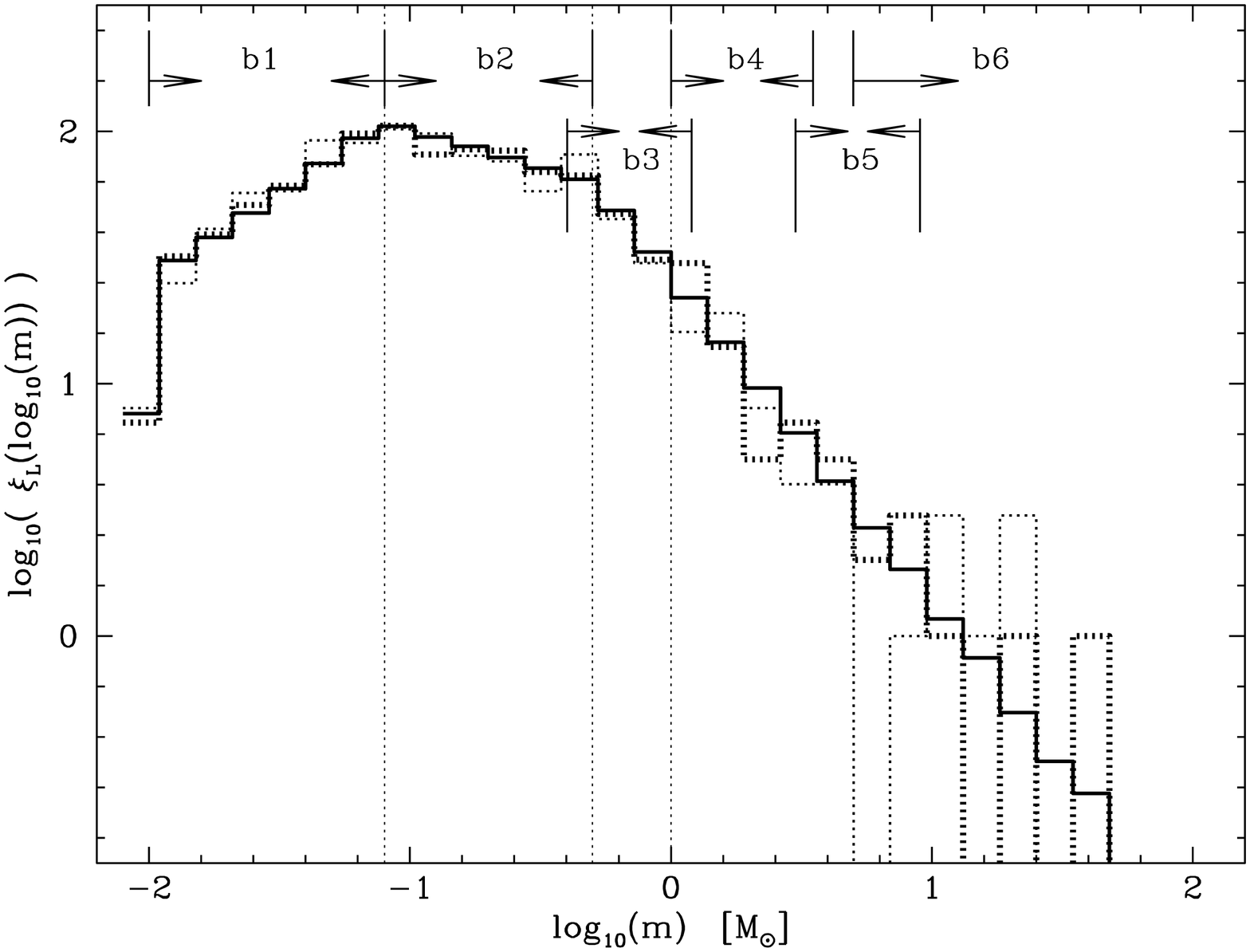}}}
\vskip 0mm
\caption
{\small{ The adopted logarithmic IMF (eqs.~\ref{eq:imf}
and~\ref{eq:xiL}), $\xi_{\rm L}/10^3$, for $10^6$ stars (solid
histogram). Two random renditions of this IMF with $10^3$ stars are
shown as the heavy and thin dotted histograms. The mass-ranges over
which power-law functions are fitted are indicated by the arrowed six
regions (eq.~\ref{eq:bin}), while thin vertical dotted lines indicate
the masses at which $\alpha_i$ changes.  }}
\label{fig:mf_sv}
\vfill
\rotatebox{-90}{\resizebox{0.6 \textwidth}{!}
{\includegraphics{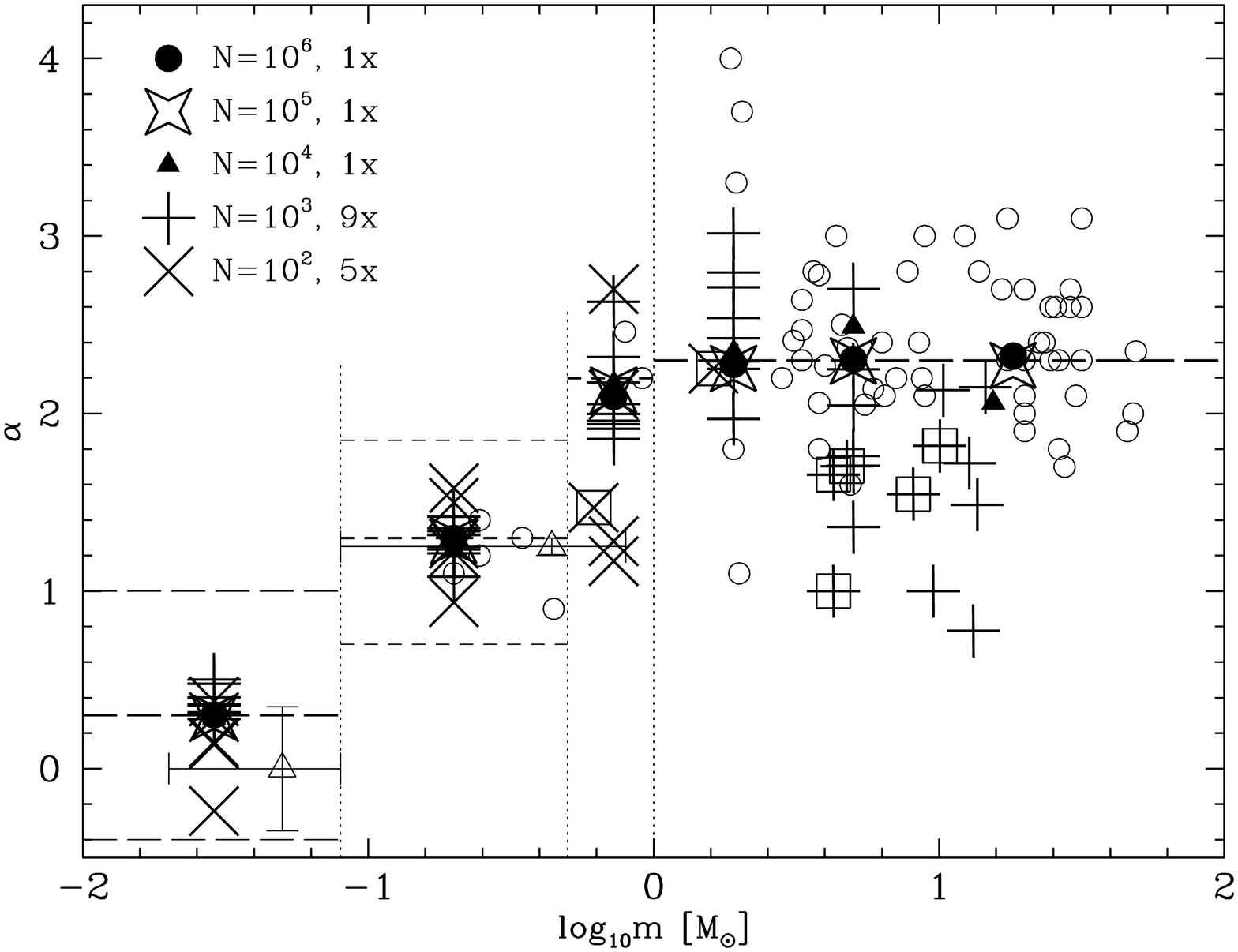}}}
\vskip 0mm
\caption
{\small{ Purely statistical variation of $\alpha$ in the six mass
ranges (eq.~\ref{eq:bin}) for different $N$ as indicated in the key.
Large outer squares indicate those $\alpha$ fits obtained with $nb=2$
and~3 mass bins.  The open circles, open triangles, vertical and
horizontal lines are as in Fig.~\ref{fig:a_m}.  }}
\label{fig:a_m_sv}
\end{center}
\end{figure}

Fig.~\ref{fig:mf_sv} illustrates this {\it sampling bias}. The
under-sampling of the histogram in the highest mass bins, when $N$ is
too small, leads to an apparent flattening of the MF in the most
massive bins accessible to the stellar population, as is evident in
Fig.~\ref{fig:a_m_sv}.  It is also evident in fig.~2 of Elmegreen
(1999), and in typical star-count data, such as used by Massey et
al. (1995b, their fig.~5) to infer the power-law index.  Such samples
contain typically a few dozen stars only (their table~5).  This is
interesting, possibly implying that the correct single-star IMF may be
steepening, i.e. have an increasing $\alpha$, with l$m$ at the largest
masses, since the uncorrected data suggest a constant $\alpha$ for
$m\simgreat1\,M_\odot$. This issue, together with the bias through the
high multiplicity fraction, will require more explicit modelling of
the biases affecting the observed IMF for massive stars.

In conclusion, Fig.~\ref{fig:a_m_sv} demonstrates that the observed
scatter is arrived at approximately for populations that contain
$10^2\simless N\simless 10^3$ stars, which is quite typical for the
type of samples available.

\section{STAR CLUSTER MODELS}
\label{sec:clmods}
\noindent 
In Section~\ref{sec:sv} apparent variations of the IMF are discussed
that result purely from statistical fluctuations.  Additional sources
of uncertainty are listed in the Introduction. Section~\ref{sec:res}
concentrates on quantifying the apparent variations that arise from
stellar-dynamical effects and unresolved binary systems. To achieve
this, a range of star-cluster models are constructed.  This approach
is relevant to populations in young clusters, OB associations and even
the Galactic field, because most stars form in embedded clusters (Lada
\& Lada 1991; Kroupa 1995b).

\subsection{Codes}
\label{sec:codes}
The dynamical evolution of the clusters studied here is calculated
using {\sc Nbody6} (Aarseth 1999; Aarseth 2000), which includes
state-of-the art stellar evolution (Hurley, Pols \& Tout 2000), a
standard Galactic tidal field (Terlevich 1987), and additional
routines for initiating the binary-rich population (Kroupa 1995c).
The $N$-body data are analysed with a large data-reduction programme
that calculates, among many quantities, the binary proportion and MFs.

\subsection{The clusters}
\label{sec:cl}
The cluster models are set-up to have the same central density,
$\rho_{\rm C}=10^{4.8}$~stars/pc$^3$, as observed in the Trapezium
Cluster (McCaughrean \& Stauffer 1994), giving a half-mass diameter
crossing time $t_{\rm cross}=0.24$~Myr.  The centre of masses of the
binary systems follow a Plummer density distribution (Aarseth, H\'enon
\& Wielen 1974) with half-mass radius $R_{0.5}$. The average stellar
mass is independent of the radial distance, $R$, from the cluster
centre, and the clusters are in virial equilibrium. Their parameters
are listed in Table~\ref{tab:clmods}. Cluster evolution is followed
for 150~Myr. 

\begin{table}
{\small
\begin{minipage}[t]{10cm}
\vskip 5mm
\begin{tabular}{cccccccc}

model &$N$ &$N_{\rm bin}$ &$R_{0.5}$ &$<\!\!m\!\!>$ 
&$\sigma_{\rm 3D}$ &$t_{\rm rel}$ &$N_{\rm run}$\\
      &    &              &[pc]      &[$M_\odot$]           
&[km/s]            &[Myr] \\
B800   &800   &400    &0.19    &0.4 &1.6 &0.8--1.4 &5\\
B3000  &3000  &1500   &0.30    &0.4 &2.5 &2.4--4.4 &5 \\
B1E4   &$10^4$ &5000  &0.45    &0.4 &3.7 &6.8--12.5     &2 \\
B1E4d  &$10^4$ &5000  &0.45    &0.3 &3.2 &7.9--14.5     &2 \\

\end{tabular}
\end{minipage}
}
\caption{\small{Cluster models: $N$ and $N_{\rm bin}$ are the initial
number of stars and binaries in each model (not taking into account
mergers), $R_{0.5}$ is the half-mass radius, and $<\!\!m\!\!>$ is the
average stellar mass. The three-dimensional velocity dispersion is
$\sigma_{\rm 3D}$, the median relaxation-time is $t_{\rm rel}$. It's
range results from assuming $f=1$ and $f=0$, respectively, since $f$
evolves.  The number of calculations per model is $N_{\rm run}$. Model
B1E4d has $\alpha_3=2.7$ ($m>1\,M_\odot$, eq.~\ref{eq:imf}), whereas
the other (default) models have $\alpha_3=\alpha_2=2.3$. It took about
4~months to assemble these data on standard desk-top computers. }}
\label{tab:clmods}
\end{table}

\subsection{The stellar population}
\label{sec:pop}
Stellar masses are distributed according to the IMF (eq.~\ref{eq:imf})
with $m_{\rm l}=0.01\,M_\odot$ and $m_{\rm u}=50\,M_\odot$. This upper
mass limit is half as large as the upper limit on the mass range used
to evaluate the MF (eq.~\ref{eq:bin}), to take into account stellar
mergers.  Merging can occur during pre-main sequence eigenevolution,
as detailed below.  The default models assume $\alpha_3=\alpha_2$, but
one model is also constructed with the possibly more realistic value
$\alpha_3=2.7$ (this model has $\alpha_0=0.5$ for historical reasons).

Binaries are created by pairing the stars randomly. The binary
proportion is
\begin{equation}
f={N_{\rm bin}\over N_{\rm sing}+N_{\rm bin}},
\label{eq:f}
\end{equation}
where $N_{\rm sing}$ and $N_{\rm bin}$ are the number of single-star
and binary systems, respectively.  A birth binary proportion $f=1$ is
assumed.  The initial mean system mass is $2<\!\!m\!\!>$, with
$<\!\!m\!\!>$ being the average stellar mass.  This results in an
approximately flat mass-ratio distribution (fig.~12 in Kroupa
1995c). Note however that encounters in clusters lead to the preferred
disruption of binaries with low-mass companions. The initially
``random'' mass-ratio distribution evolves rapidly towards a
distribution in which low-mass companions are less frequent, but still
preferred (Kroupa 1995c). This is consistent with observations in that
G-dwarf primaries (Duquennoy \& Mayor 1991), Cepheids ($4-9\,M_\odot$,
Evans 1995) and possibly OB stars (Mason et al. 1998; Preibisch et
al. 1999) prefer low-mass companions.

Periods and eccentricities are distributed following Kroupa (1995c).
The periods range from about 1~d to $10^9$~d, and {\it pre-main
sequence eigenevolution} changes the periods, mass ratios and
eccentricities such that they are consistent with observational
constraints for late-type main sequence stars with short
periods. Eigenevolution is the collective name for system-internal
processes that evolve the orbital parameters, such as tidal
circularisation, mass transfer, and interactions with circum-stellar
and circum-binary discs. One feature of the pre-main sequence
eigenevolution model is that secondary companions gain mass during
accretion if the peri-astron distance is smaller than a critical
value. This affects the IMF by slightly reducing the number of
low-mass stars, and slightly increasing the number of massive
stars. Also, in some rare cases the binary companions merge giving
$N_{\rm sing}>0$, so that the true initial binary proportion is less
than unity.  Since only short-period binaries are affected by
eigenevolution, the overall changes to the IMF are not significant.

The resulting single-star and system MFs are shown in
Fig.~\ref{fig:mf_eg}. This figure demonstrates that the IMF that
results from the eigenevolution model has a slightly smaller $\alpha$,
especially for $m<0.5\,M_\odot$ (thick solid histogram). This effect
is larger for the default case ($\alpha_3=2.3$), because the larger
number of massive stars implies more systems in which the secondary
gains mass as a result of eigenevolution. The effect on $\alpha$ is
too small, however, to make a significant difference in the alpha-plot
(e.g. Fig.~\ref{fig:a_m_N1E4} below).  Fig.~\ref{fig:mf_eg} also
displays the large difference between the system MF and the
single-star MF at low masses. The IMF has a maximum near
$0.1\,M_\odot$, whereas the system MF has one near $0.4\,M_\odot$, and
underestimates the number of 'stars' by an order of magnitude near
$m=0.01\,M_\odot$, and by a factor of three near $m=0.08\,M_\odot$.
\begin{figure}
\begin{center}
\rotatebox{-90}{\resizebox{0.5 \textwidth}{!}
{\includegraphics{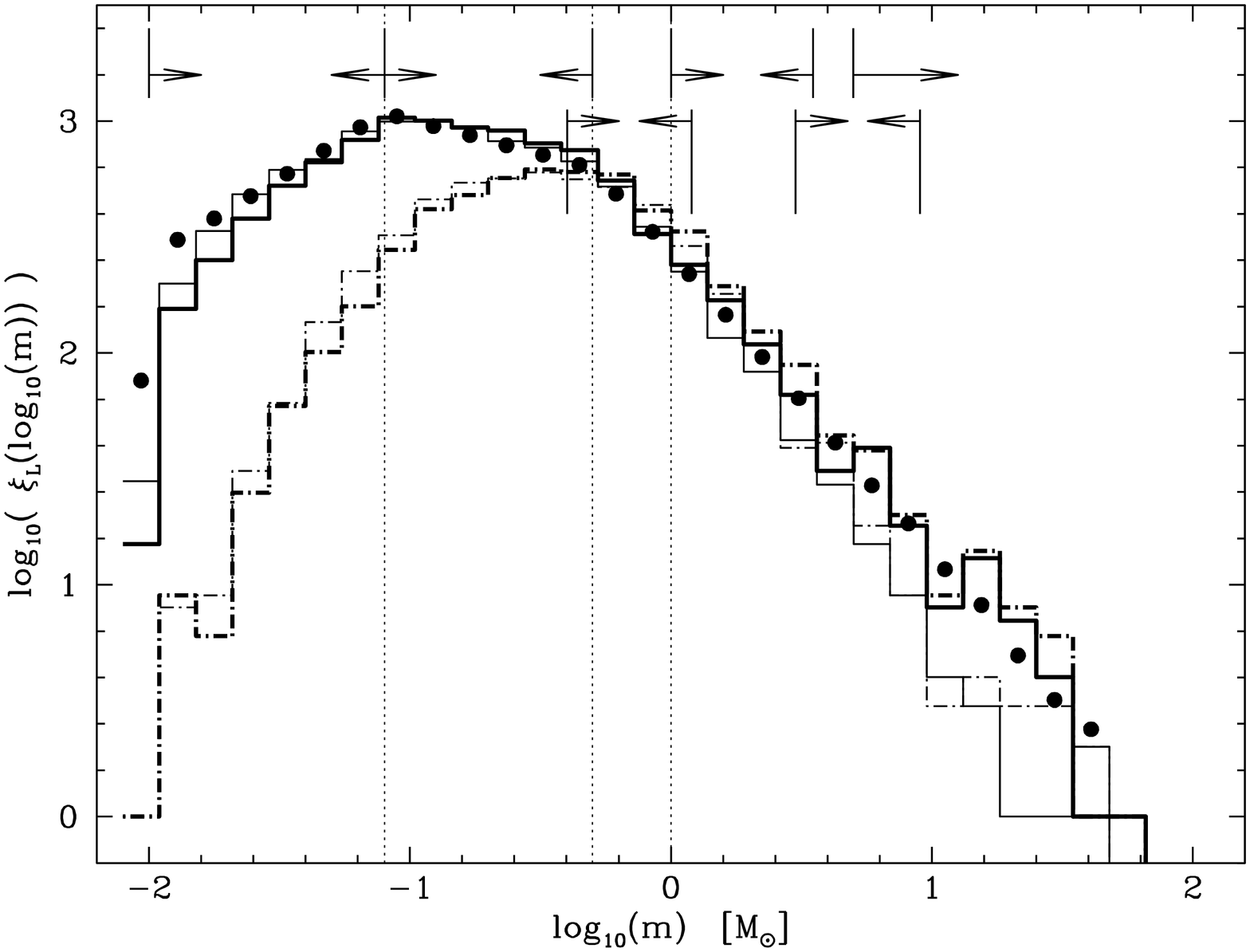}}}
\vskip 0mm
\caption
{\small{ 
Mass functions for single-stars (solid histograms) and
systems (dash-dotted histograms) at $t=0$ in models~B1E4 (thick
histograms) and~B1E4d (thin histograms). Note the smaller number of
massive stars in model~B1E4d, which has a steeper IMF for
$m>1\,M_\odot$ with $\alpha_3=2.7$ (Table~\ref{tab:clmods}). The solid
dots are the IMF for $N=10^6$ stars (Fig.~\ref{fig:mf_sv}) scaled to
$N=10^4$, and the vertical dotted lines and arrowed regions are as in
Fig.~\ref{fig:mf_sv}.
}}
\label{fig:mf_eg}
\end{center}
\end{figure}

\subsection{Nota bene}
\label{sec:nb}
The {\it cluster models constructed here are extremes}, in that they
have a very high central density equal to that observed in the
ONC. This assumption leads to a {\it rapid depletion of the binary
population}, as shown below (Fig.~\ref{fig:fpm}; see also de La Fuente
Marcos 1997). Disruption of binaries occurs on a crossing-time scale
(Kroupa 2000a) in any cluster, so that it takes much longer in real
time for the binary population to decrease in a Pleiades-type cluster,
for which K\"ahler (1999) shows that $f\approx0.7$ is
possible. Likewise, the pre-main-sequence cluster IC348, which has a
density of about 500~stars/pc$^3$, has a binary proportion similar to
that in the Galactic field (Duchene, Bouvier \& Simon 1999). As shown
by KTG91, such a binary proportion requires significant correction to
the observed system LF to infer the IMF. The problem with unresolved
binaries may be even worse for lower-density clusters still, such as
studied by Testi, Palla \& Natta (1999), because the binary population
evolves on much longer time-scales, and is thus likely to be less
evolved than in the clusters studied here. The problem will {\it
never} be smaller in such clusters, unless they consist of a stellar
population that had an unusually small initial binary proportion
($f<0.3$), i.e. smaller than even in the evolved extreme models
here. Such a population has never been observed in any Galactic
cluster or association to this date (e.g. Ghez et al. 1997; Duchene
1999).

Any real population is thus likely to have a {\it larger binary
proportion than in the models considered here} after about three
crossing times ($\approx0.8$~Myr). In addition, the present results
will be an underestimate of the bias, because only binary systems are
considered. Real populations contain something like 20~per~cent or
more triple and quadruple systems, which, when not resolved, increase
the systematic error made in the observational estimate of $\alpha$.
What is inferred in this paper is thus the {\bf\it minimum correction
to} $\mathbf{\alpha}$.

This is particularly true for $m\simgreat1\,M_\odot$, because the
observed mass-ratio distribution for massive stars (e.g. Preibisch et
al. 1999) has secondaries that are typically more massive than
$1\,M_\odot$, whereas in the models here, massive primaries typically
have very low-mass companions owing to the random sampling
hypothesis. This is very important when considering the system MFs
below. It will be evident that the models lead to essentially no bias
for massive stars, but this is more likely to be a shortcoming of the
present assumptions, rather than proving that the IMF for massive
systems is not subject to a significant bias, as discussed in
Section~\ref{sec:ms}. Clearly, this is a fundamentally important topic
requiring much more work to construct a more realistic initial
mass-ratio distribution for massive stars. In addition, a
systematically different IMF between the LMC and the MW for massive
stars may become evident, {\it if} the binary properties differ
systematically between the two galaxies, because then the correction
for systematic bias would be different for the two samples. At present
no such difference is known, and so the empirical LMC and MW data
plotted in Fig.~\ref{fig:a_m} can, at present, be only taken to mean
that the IMF for massive stars may be the same in the two populations.

\section{RESULTS}
\label{sec:res}
\noindent 
The results obtained from the stellar-dynamical calculations are used
to study temporal and spatial apparent variations of the single-star
and system MFs.

\subsection{Cluster evolution}
\label{sec:clev}
\noindent 
As an impression of the evolution of the star clusters,
Fig.~\ref{fig:nfev} displays the scaled number of {\it systems} and
single stars with $R\le3.2$~pc.  $N_{\rm sys}(t)=N_{\rm
sing}(t)+N_{\rm bin}(t)$ increases for $t\simless2.5$~Myr because the
disruption of binary systems liberates secondaries. That is, the
observer would find that the number of 'stars' increases with time.
After $t\approx2.5$~Myr, $N_{\rm sys}$ decreases with a rate depending
on $N$, because the clusters expand owing to binary-star heating,
relaxation and mass-loss from evolving stars.
\begin{figure}
\begin{center}
\rotatebox{0}{\resizebox{0.77 \textwidth}{!}
{\includegraphics{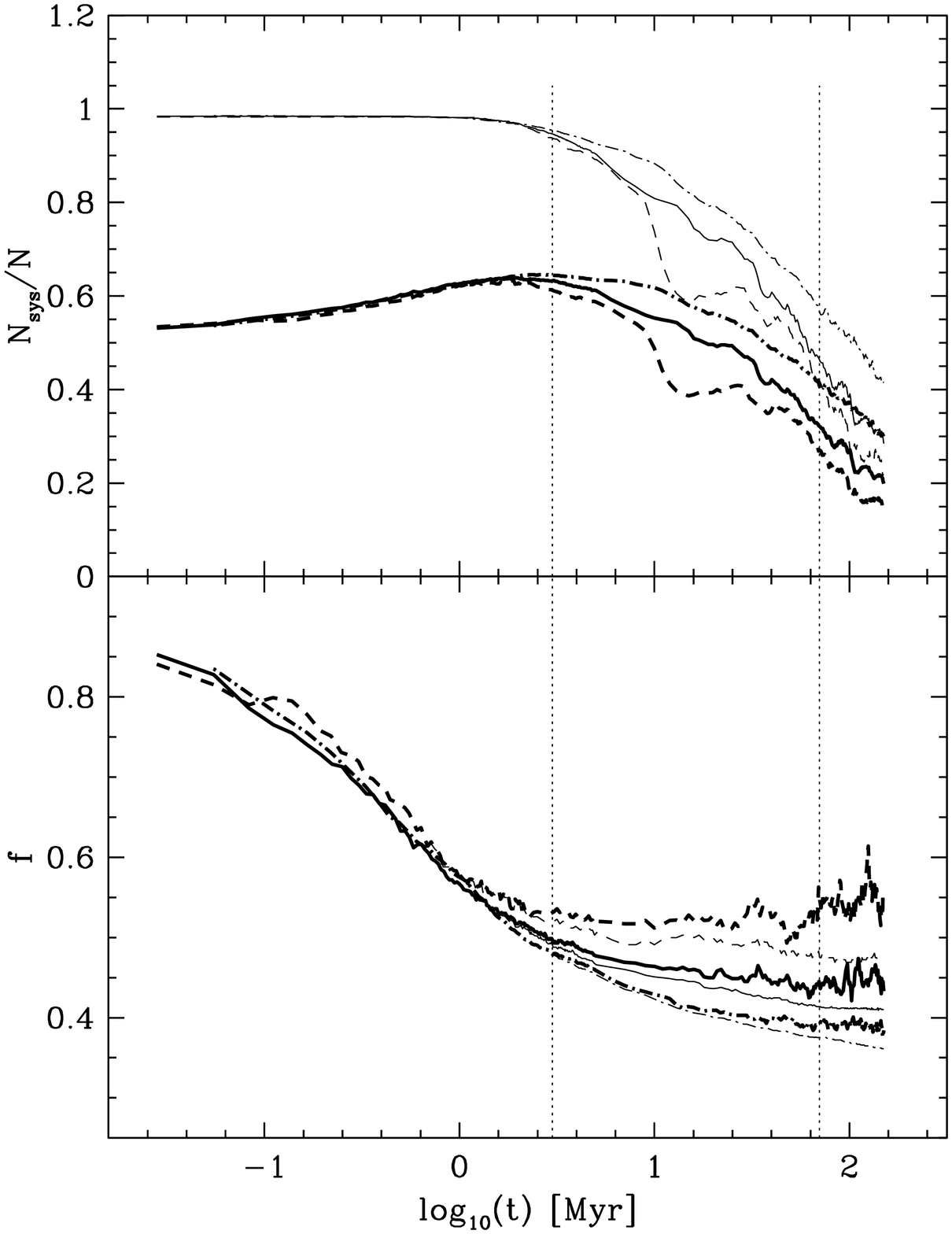}}}
\vskip 0mm
\caption
{\small{ 
Examples of the evolution of individual clusters.  {\bf Top
panel:} The number of systems (thick curves) and all individual stars
and BDs (thin curves) within the innermost 3.2~pc.  The short-dashed
lines are for $N=800$, the solid lines are for $N=3000$, and the
dash-dotted lines are for $N=10^4$. {\bf Bottom panel:} The binary
proportion for $R\le3.2$~pc (thick curves), and all $R$ (thin curves)
for the same cases as in top panel.  In both panels, the horizontal
dotted lines indicate the times (3~and 70~Myr) at which the mass
functions are observed.}}
\label{fig:nfev}
\end{center}
\end{figure}

The binary proportion (Fig.~{\ref{fig:nfev}) decreases within a few
initial crossing times. The decay occurs on exactly the same
time-scale for the different clusters, demonstrating that it is not
the velocity dispersion in the cluster alone which dictates the
disruptions, but the density as well.  Owing to the ejection from the
cluster of preferably single stars and because of mass segregation,
$f$ is larger for systems with $R\le3.2$~pc and at times
$t\simgreat2.5$~Myr, than for systems at larger distances from the
clusters. The least massive clusters ($N=800$) have expanded
appreciably by this time so that the remaining binary population in
the cluster is hard, and no further significant disruption of binaries
occurs ($f\approx0.55$ and increasing for $t\simgreat2.5$~Myr). The
more massive clusters, however, remain more concentrated for a longer
time (top panel of Fig.~\ref{fig:nfev}), and consequently the
binary-star hard/soft boundary remains at a higher binary binding
energy for a longer time. At any time $t\simgreat 3\,t_{\rm
cross}\approx 0.7$~Myr, the binary proportion is higher in the
clusters with smaller $N$, which his particularly evident in
Fig.~\ref{fig:fpm} below. This is a nice example of the caveat raised
in Section~\ref{sec:nb}.  Further details on these processes are
available in Kroupa (2000b), and in the seminal paper by Heggie
(1975).

The evolution of the binary proportion for primaries with different
masses is illustrated in Fig.~\ref{fig:fpm}. The binary proportion of
BDs falls rapidly, and stabilises near $f_{\rm BD}=0.20$ in all
models. M~dwarfs retain a much higher binary proportion by
$t=150$~Myr, $f_{\rm M}\approx0.4-0.5$, depending on $N$, and more
massive primaries retain a slightly higher binary proportion
still. The overall binary proportion of O~primaries ($m\ge8\,M_\odot$)
shows a complex behaviour. Initially, most O primaries have low-mass
companions. These are, however, exchanged for more massive companions
near the cluster core. When the primaries explode, these companions
are left or are ejected as single stars.  In addition, violent
dynamical encounters in the cluster core eject single massive stars.
Overall, $f_{\rm O}$ decays, but higher-order multiplicities that form
in three-body encounters are not accounted for.
\begin{figure}
\begin{center}
\rotatebox{0}{\resizebox{0.77 \textwidth}{!}
{\includegraphics{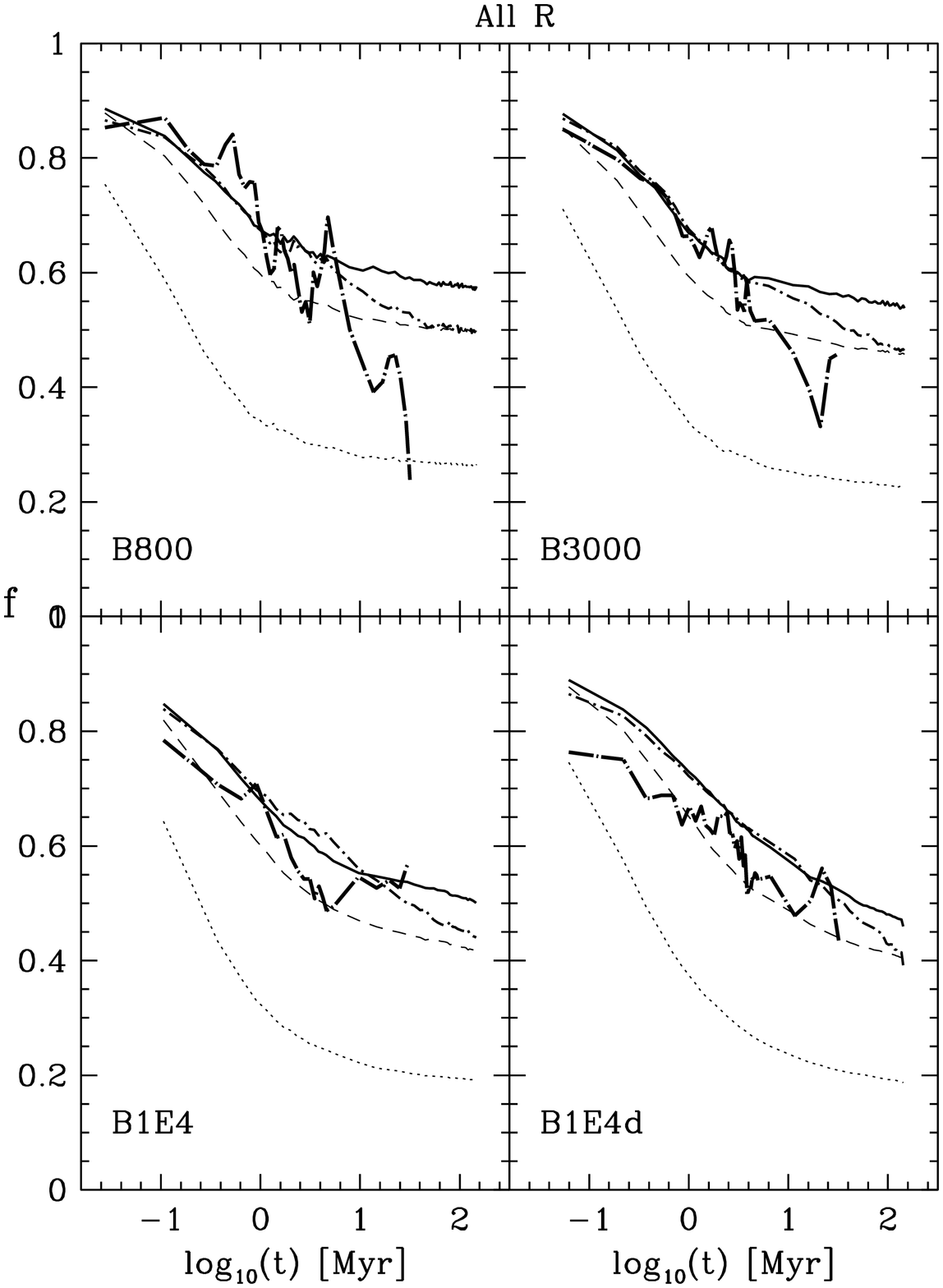}}}
\vskip 0mm
\caption
{\small{ 
The proportion of binaries with primary masses
$>8\,M_\odot$, $f_{\rm O}$ (thick long-dash-dotted curve),
$1-8\,M_\odot$, $f_{\rm IM}$ (thin short-dash-dotted curve), and
$0.5-1\,M_\odot$, $f_{\rm K}$ (solid curve). M~dwarf primaries
($0.08-0.5\,M_\odot$) have a binary proportion, $f_{\rm M}$ (thin
dashed line), whereas brown dwarfs ($0.01-0.08\,M_\odot$), $f_{\rm
BD}$, are shown as the thin dotted line. Each curve is an ensemble
mean.
}}
\label{fig:fpm}
\end{center}
\end{figure}

\subsection{The alpha plot for cluster populations}
\label{sec:amclpop}
\noindent 
Having briefly discussed the evolution of the clusters and of the
binary population, the following question can now be posed: What MFs
would an observer deduce if an ensemble of such clusters were observed
at different times, under the extreme assumption that the mass of each
star or system can be measured exactly?

Figs.~\ref{fig:a_m_N800} to~\ref{fig:a_m_N1E4} show the results for
each $N$. The upper panels assume the observer sees all stars with
$R\le3.2$~pc, whereas in the lower panel it is assumed that only the
system masses can be measured exactly for systems with
$R\le3.2$~pc. The MFs are constructed at times $t=0$, 3~Myr and
70~Myr. For the single-star MFs, the results at $t=0$ are the same as
for pure statistical noise (Fig.~\ref{fig:a_m_sv}).
\begin{sidewaysfigure}
\begin{center}
\rotatebox{-90}{\resizebox{0.77 \textwidth}{!}
{\includegraphics{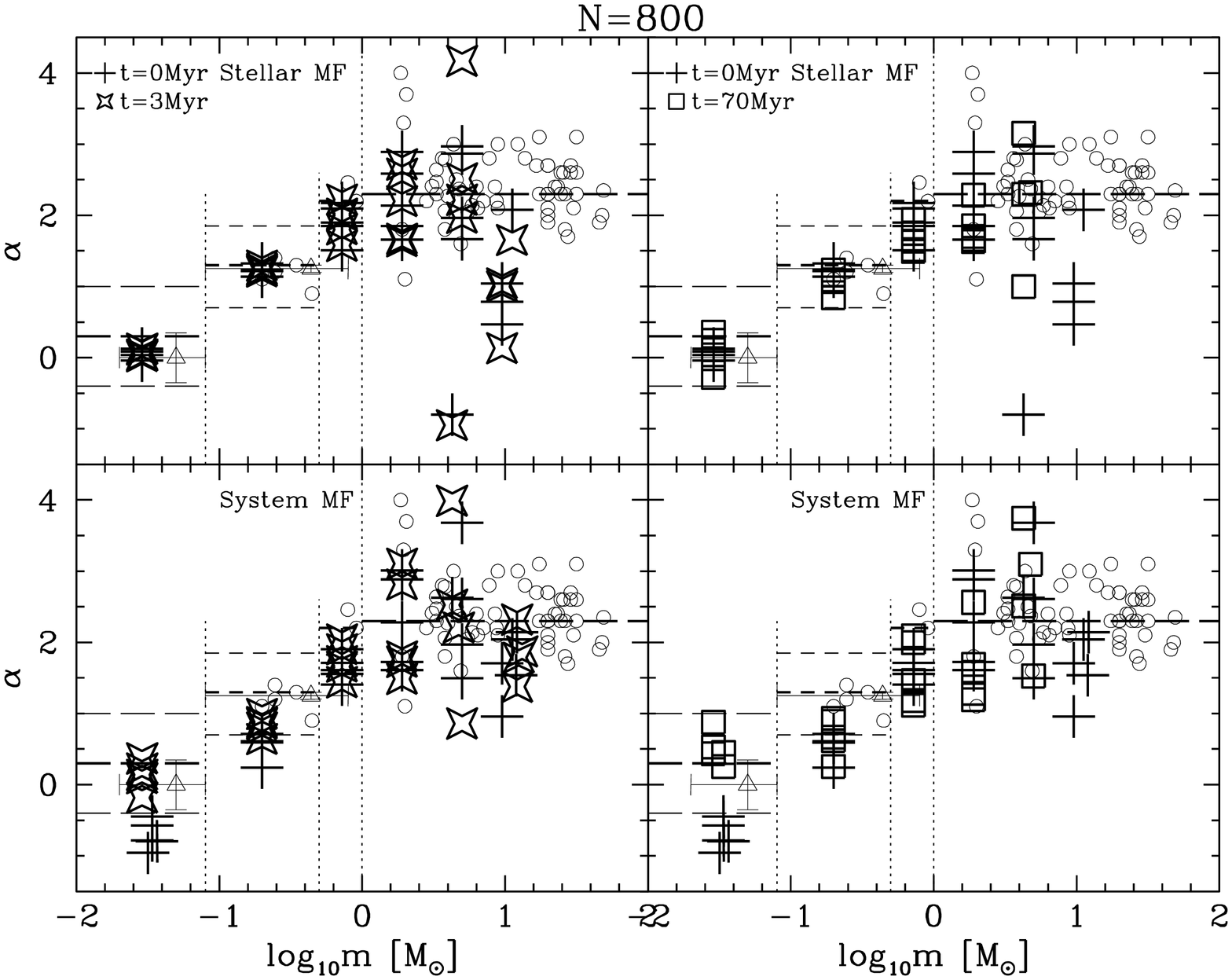}}}
\vskip 0mm
\caption
{\small{ The $\alpha$-plot for five~B800 models at $t=0$, 3~Myr (left
panels) and 70~Myr (right panels). The single-star (top panels) and
system (lower panels) MFs are constructed for stars with
$R\le3.2$~pc. The open circles, open triangles, vertical and
horizontal lines are as in Fig.~\ref{fig:a_m_sv}.  }}
\label{fig:a_m_N800}
\end{center}
\end{sidewaysfigure}

\begin{sidewaysfigure}
\begin{center}
\rotatebox{-90}{\resizebox{0.77 \textwidth}{!}
{\includegraphics{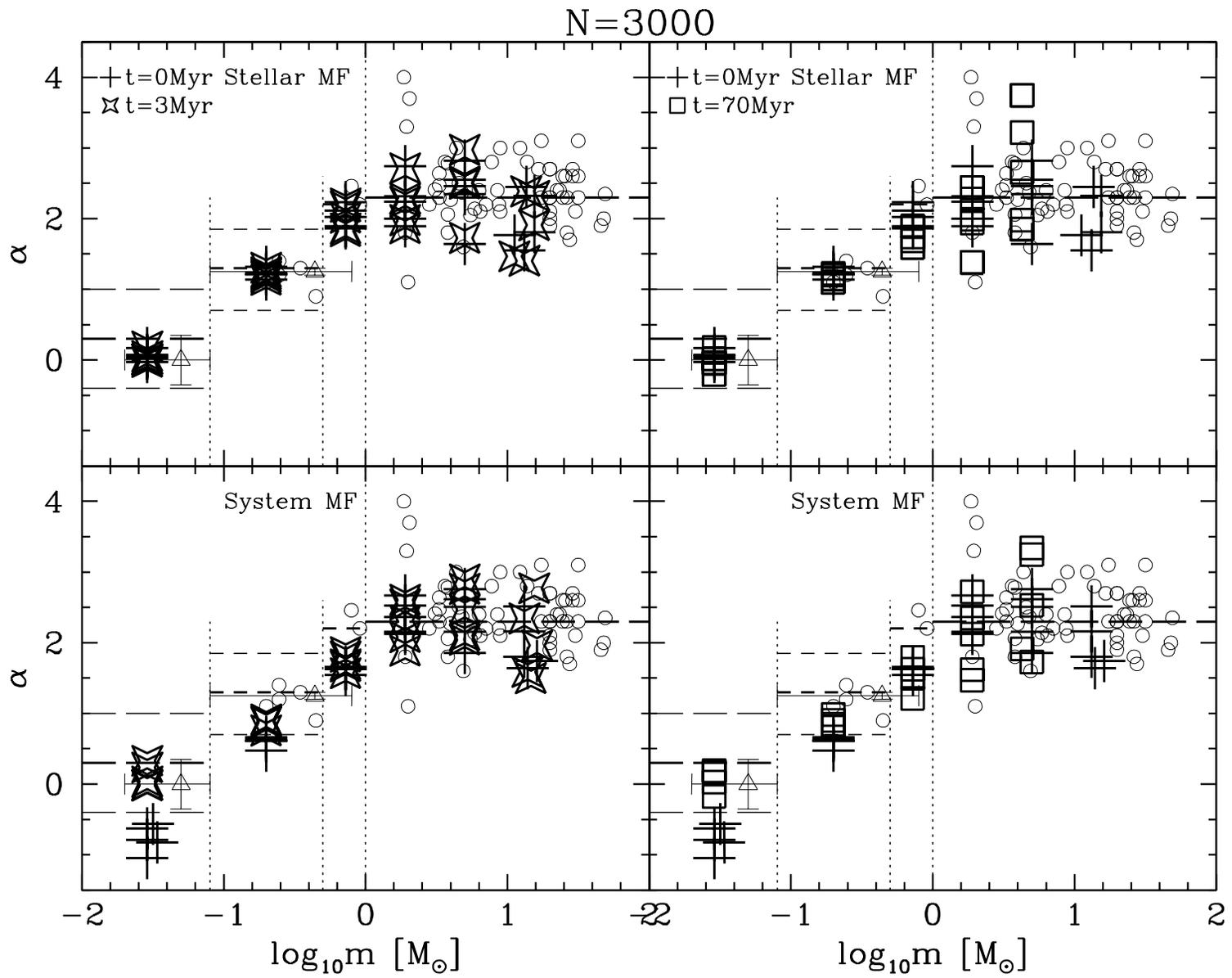}}}
\vskip 0mm
\caption
{\small{ 
As Fig.~\ref{fig:a_m_N800}, but for five~B3000 models.}}
\label{fig:a_m_N3000}
\end{center}
\end{sidewaysfigure}

\begin{sidewaysfigure}
\begin{center}
\rotatebox{-90}{\resizebox{0.77 \textwidth}{!}
{\includegraphics{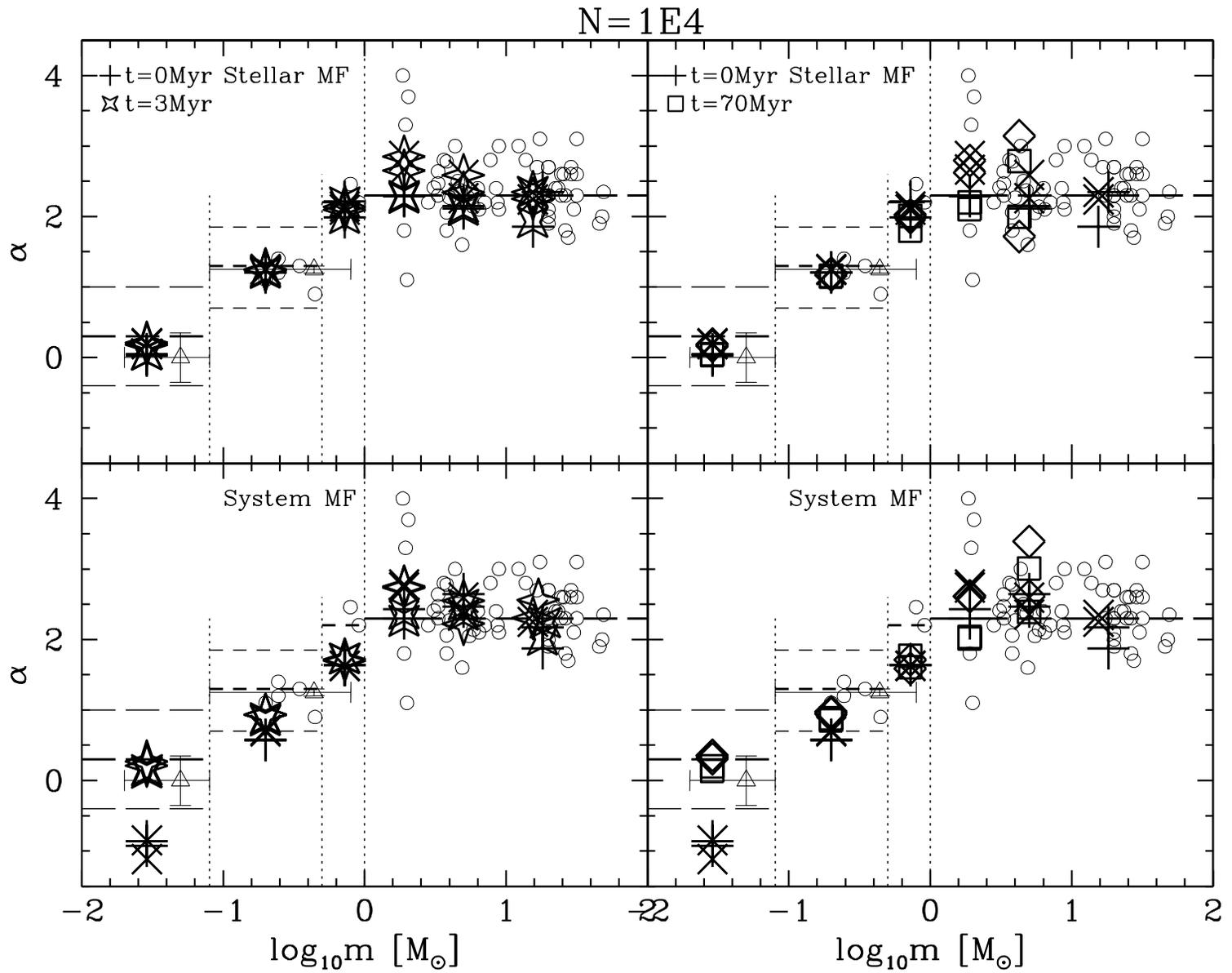}}}
\vskip 0mm
\caption
{\small{ As Fig.~\ref{fig:a_m_N800}, but for models with $N=10^4$
stars.  The crosses, four-pointed stars and squares are for the two
B1E4 models, whereas the same symbols but rotated by 45 degrees are
for the two B1E4d models. }}
\label{fig:a_m_N1E4}
\end{center}
\end{sidewaysfigure}

At $t=0$, the single-star IMF is well reproduced.  The system MF, on
the other hand, underestimates $\alpha$ significantly for $m_{\rm
sys}<1\,M_\odot$, with $\alpha_0\approx-0.8$ (instead of $+0.3$) for
$m\simless0.08\,M_\odot$, $\alpha_1\approx+0.7$ (instead of $+1.3$)
for $0.08\simless m\simless0.5\,M_\odot$, and $\alpha_2\approx+1.5$
(instead of $+2.3$) for $0.5\simless m \simless 1\,M_\odot$.

At $t=3$~Myr and~70~Myr, most of the BD systems have been disrupted
(Fig.~\ref{fig:fpm}), with typically $f_{\rm BD}\approx0.2$, and most
star--BD systems have also ceased to exist, so that $\alpha_0$ is only
slightly underestimated for the system MF. Work is in progress to
study if the resulting mass-ratio distribution becomes consistent with
the observed 'BD-companion dessert' for nearby stars (M. Mayor 2000,
priv.commun.).  In mass ranges $b2$ and~$b3$, the power-law index is
still underestimated significantly, because the surviving binary
proportion is typically $f>0.4$ for $m>0.08\,M_\odot$.  For $b2$ the
lower panels in Figs.~\ref{fig:a_m_N800} to~\ref{fig:a_m_N1E4} read
$\alpha_1\approx+0.8$, and in $b3$, $\alpha_2\approx+1.7$. The bias in
measuring $\alpha_{1,2}$ for the system MF rather than the single-star
MF is thus not significantly reduced at later times.

This bias will operate for even older clusters, because further binary
disruption is essentially halted in the expanded clusters, and $f$
begins to increase with time as energy equipartition retains the
heavier binaries in the cluster at the expense of single stars (fig.~3
in Kroupa 1995d).  However, with time the bias will decrease for
$\alpha_2$ as the turnoff mass becomes smaller, i.e. as the number of
primaries with $m\simgreat1\,M_\odot$ decreases.  As an extreme
example, globular clusters retain a significant proportion of their
low-mass stars (Vesperini \& Heggie 1997), but stars with
$m\simgreat0.8\,M_\odot$ have ceased to exist, so that no
$m\simless0.8\,M_\odot$ companions are 'hidden' with brighter
primaries.

For $N=800$ (Fig.~\ref{fig:a_m_N800}) the scatter in range $b5$ is
very large, and rather similar to what is seen in the observational
data in the shaded area ($0.8-3\,M_\odot$, Fig.~\ref{fig:a_m}). This
is interesting, because in these models it is the stars in the mass
range $3-9\,M_\odot$ that are the most massive {\it and} abundant
enough to eject each other from the core after meeting there through
mass segregation, causing large fluctuations in the measured MF.  The
same holds true for the cluster data in the shaded range in
Fig.~\ref{fig:a_m}. For example, $\rho$~Oph contains not more than a
few hundred systems, so that the most massive stars populate roughly
the shaded range. The Pleiades is 100~Myr old, so that stars with
$m\simgreat10\,M_\odot$ have evolved off the main sequence, and the
stars just below this mass interact in the cluster core.  

In summary, comparison of the three figures shows that the scatter in
$\alpha$ decreases as $N$ increases, but that the scatter is larger
than pure Poisson noise (cf. the $t>0$ data in the upper panel of
Fig.~\ref{fig:a_m_N1E4} with the $N=10^4$ model in
Fig.~\ref{fig:a_m_sv}). The most important result though is that
$\alpha_{1,2}$ is underestimated by $\Delta\alpha\approx0.5$ for the
system MF in the mass range $0.1-1\,M_\odot$. And, an observer deduces
fewer BDs in an unevolved population ($t=0$; Fig.~\ref{fig:mf_eg})
such as in Taurus--Auriga, than in a population that is older than a
few crossing times, such as the Trapezium and the Pleiades Clusters
(see also Kroupa, Aarseth \& Hurley 2001).  The figures also show that
for a single-age population the scatter is always smaller for
$m\simless1\,M_\odot$. For $m\simgreat1\,M_\odot$ the scatter for the
clusters with $N=3000$ and $N=10^4$ stars is comparable to the
observed scatter. Even when $N=10^4$, models with $\alpha_3=2.3$
cannot be differentiated from models with $\alpha_3=2.7$ in mass
ranges $b5$ and $b6$.

\subsection{The alpha plot for cluster halo populations}
\label{sec:amejpop}
\noindent 
The MF ``in'' young rich clusters can often only be determined by
avoiding the crowded central regions. This can cause systematic
uncertainties because stellar encounters lead to preferentially
lower-mass stars and preferentially single stars populating an
extended halo, or being ejected from the cluster.

The clusters with $N=3000$ and $N=10^4$ stars are used to investigate
the MF for systems lying at a distance $R>3.2$~pc from the cluster
centre.  The results are shown in Fig.~\ref{fig:a_m_out}, assuming the
observer can only determine the masses of systems. 

\begin{figure}
\begin{center}
\rotatebox{0}{\resizebox{0.77 \textwidth}{!}
{\includegraphics{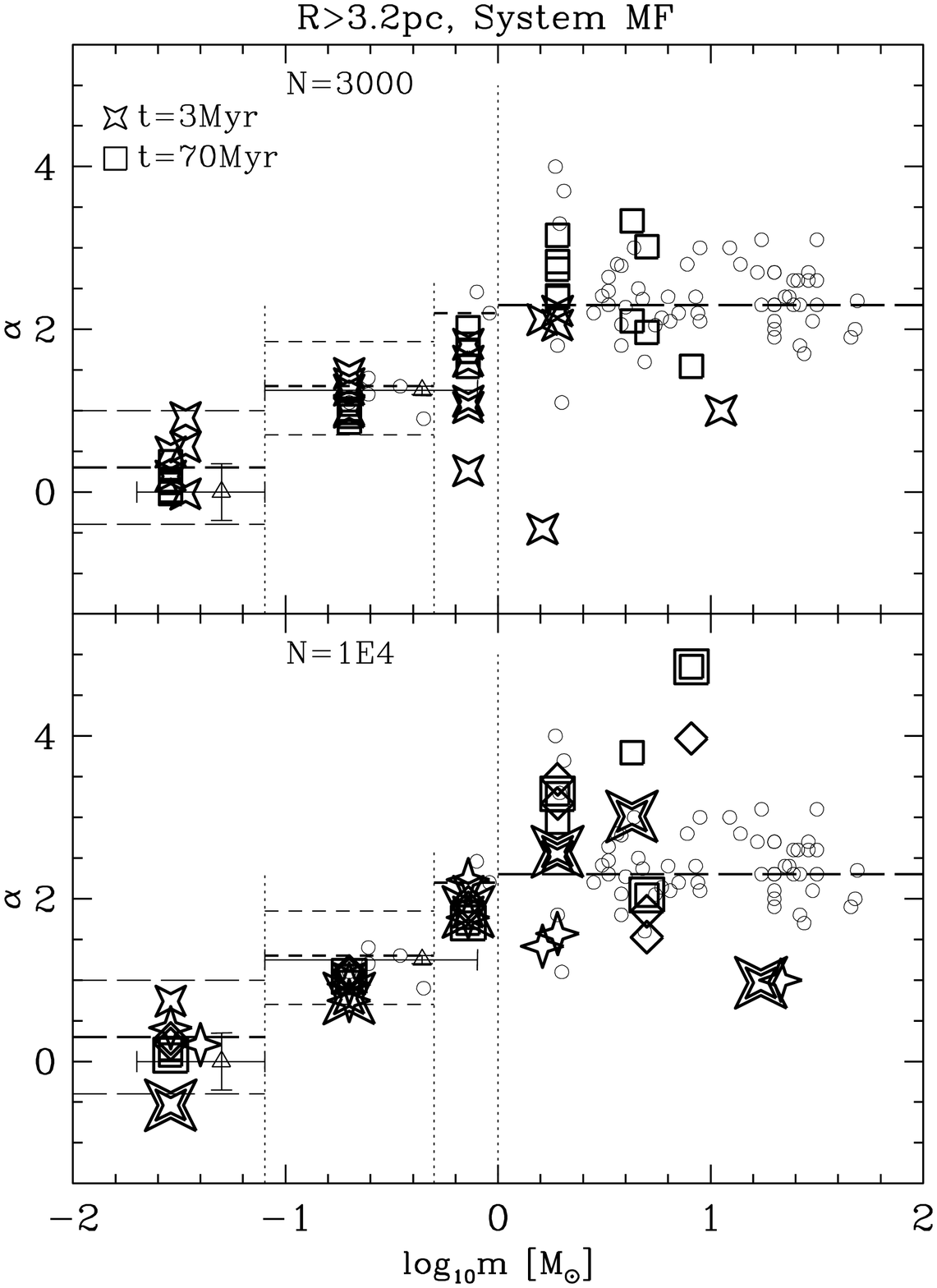}}}
\vskip 0mm
\caption
{\small{ 
Spatial variation of the MF: The MFs in models B3000 (5
renditions), B1E4 (2 renditions) and B1E4d (2 renditions) for {\it
systems} with $R>3.2$~pc. Two particularly exotic examples are
highlighted using double symbols, and the corresponding MFs are plotted
in Fig.~\ref{fig:mf_ex}, and the MF-fits are listed in
Table~\ref{tab:lm_a_eg}.  Otherwise as Fig.~\ref{fig:a_m_N1E4}.  Note
the changed $\alpha$-scale.
}}
\label{fig:a_m_out}
\end{center}
\end{figure}

The scatter is larger than within the clusters ($R\le3.2$~pc,
Section~\ref{sec:amclpop}), and the bias for $m<0.5\,M_\odot$ that
leads to an underestimate of $\alpha_1$ in binary-rich populations is
reduced significantly. This results because the halo population is
depleted in binary stars (Fig.~\ref{fig:nfev}).

Two extreme examples are marked with double symbols. The
corresponding MFs are plotted in Fig.~\ref{fig:mf_ex}. One example is
the system MF for a halo population at an age of 3~Myr. It's
particularly flat MF for $m>10\,M_\odot$ ($\alpha=0.97$) comes about
because the cluster core just expelled a few massive stars to the
outer regions.  The steep MF for a 70~Myr old population with
$\alpha=4.85$ at ${\rm l}m_{\rm av}=0.9$ (double square in the lower
panel) arises because stellar evolution has removed stars with
$m\simgreat 10\,M_\odot$, and because the stars with a mass just below
the turn-off mass are located preferably near the cluster core. The
fitted power-law indices are listed in Table~\ref{tab:lm_a_eg}.

\begin{figure}
\begin{center}
\rotatebox{-90}{\resizebox{0.6 \textwidth}{!}
{\includegraphics{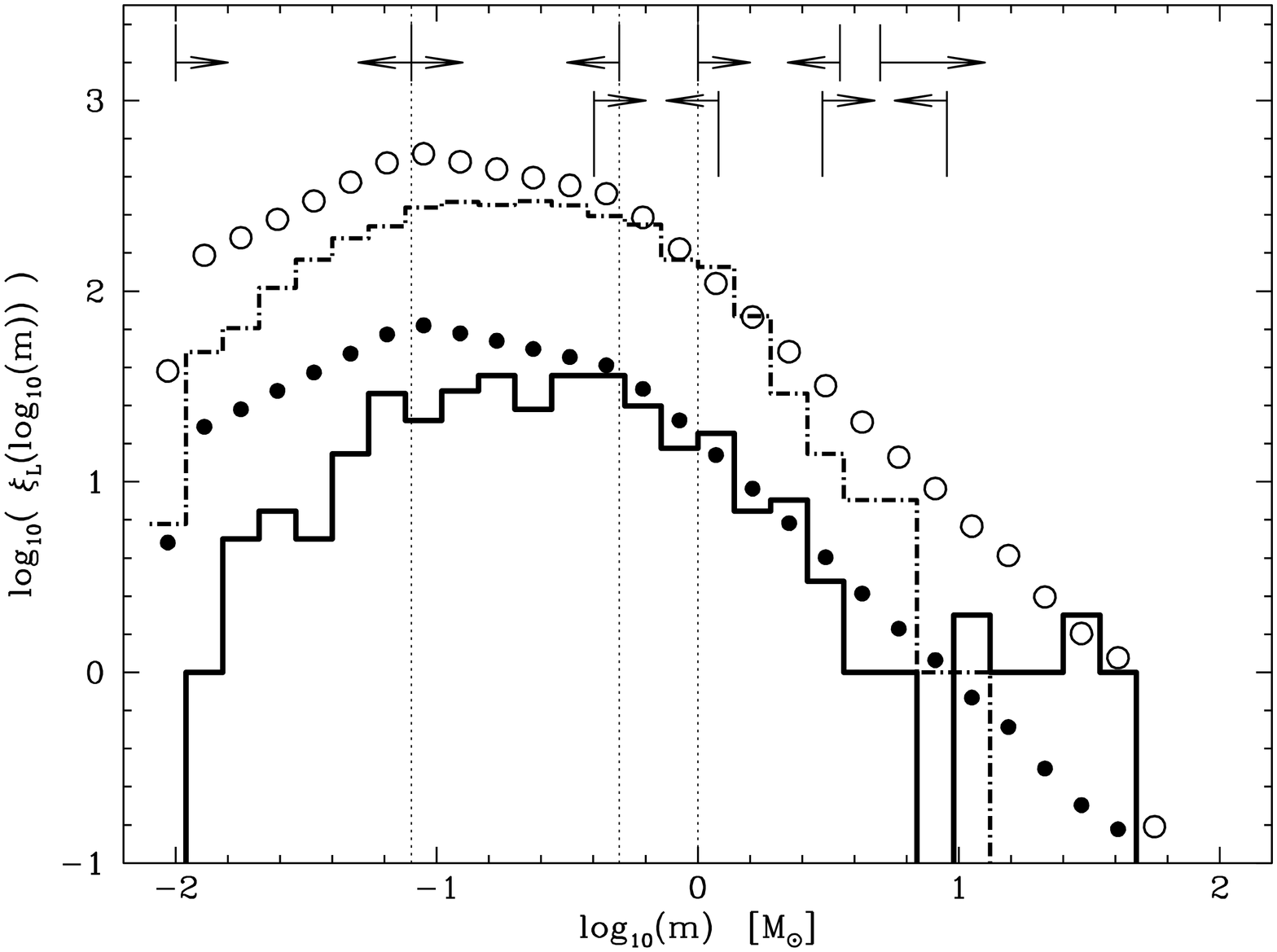}}}
\vskip 0mm
\caption
{\small{ 
Spatial variation of the MF: The MFs for systems with
$R>3.2$~pc for model B1E4 showing two cases: solid histogram,
$t=3$~Myr (four-pointed double star in Fig.~\ref{fig:a_m_out}) and
dash-dotted histogram, $t=70$~Myr (double square in
Fig.~\ref{fig:a_m_out}). The open and filled circles represent the
$N=10^6$ star IMF from Fig.~\ref{fig:mf_sv} after appropriate scaling.
The arrowed mass-ranges are as in Fig.~\ref{fig:mf_sv}.
}}
\label{fig:mf_ex}
\end{center}
\end{figure}
\vfill

\begin{table}
{\small
\begin{minipage}[b]{5cm}
\hspace{0cm}
  \begin{tabular}{*{4}{c}{l}}
      $nb$  &log$_{10}m_{\rm av}$ &$\alpha$  &$\sigma_{{\rm f},\alpha}$ \\
           &[$M_\odot$]\\
  &double star ($t=3$~Myr) in Fig.~\ref{fig:a_m_out}\\
  6   &$-1.540$ & $-0.54$    &0.31\\
  6   &$-0.700$ & $+0.77$    &0.14\\
  4   &$-0.140$ & $+1.84$    &0.29\\
  4   &$+0.280$ & $+2.58$    &0.54\\
  3   &$+0.630$ & $+3.01$    &3.42\\
  6   &$+1.237$ & $+0.97$    &1.01\\

  &double square ($t=70$~Myr) in Fig.~\ref{fig:a_m_out}\\
  6   &$-1.540$ &$+0.08$    &0.07\\
  6   &$-0.700$ &$+1.05$    &0.05\\
  4   &$-0.140$ &$+1.70$    &0.10\\
  4   &$+0.280$ &$+3.29$    &0.23\\
  4   &$+0.700$ &$+2.05$    &0.69\\
  3   &$+0.910$ &$+4.85$    &3.27\\
\end{tabular}
\end{minipage}
}
\caption{\small{The two examples highlighted in Fig.~\ref{fig:a_m_out}.  The
corresponding MFs are plotted in Fig.~\ref{fig:mf_ex}. The table lists
the number of log-mass bins used in the fit, $nb$, the average
log-mass over which the fit is obtained, ${\rm log}_{10}m_{\rm av}$,
the fitted power-law index $\alpha$ and the probable uncertainty
$\sigma_{{\rm f},\alpha}$.  }}
\label{tab:lm_a_eg}
\end{table}

\subsection{A synthetic alpha plot}
\label{sec:amsyn}
\noindent 
The results from all cluster models at different times and for 
the inner and outer cluster regions can be combined to form a
synthetic ensemble of populations. The result is shown in
Fig.~\ref{fig:a_m_allst} for the case that the observer is able to
measure the mass of each star exactly. Fig.~\ref{fig:a_m_allsys} shows
the results assuming the observer can measure the system masses
exactly. 
\begin{figure}
\begin{center}
\rotatebox{-90}{\resizebox{0.6 \textwidth}{!}
{\includegraphics{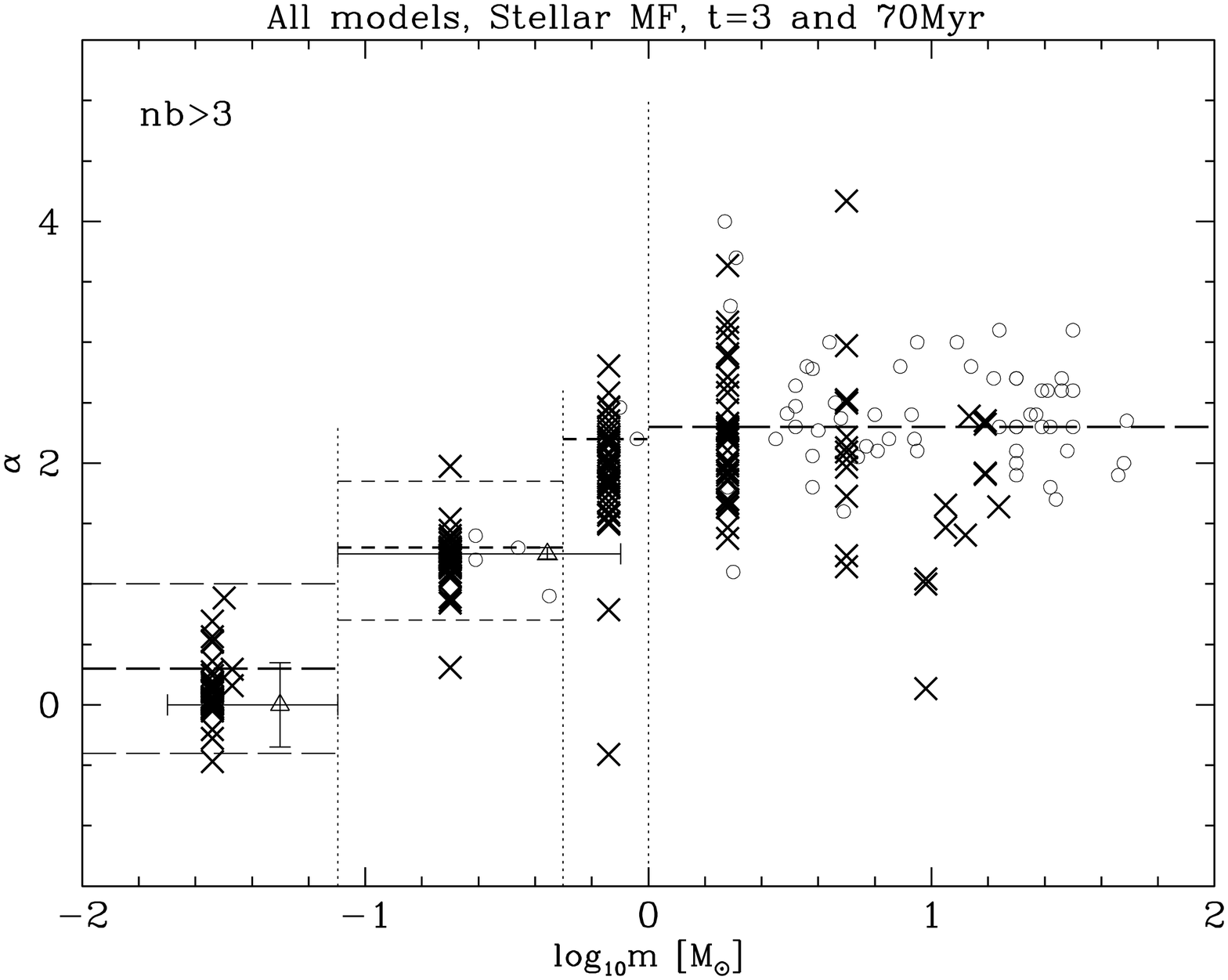}}}
\vskip 0mm
\caption
{\small{ All models B800 (5 renditions), B3000 (5 renditions) and B1E4
(2 renditions) for $t=3$~Myr and 70~Myr for {\it individual stars}
with $R\le3.2$~pc and $R>3.2$~pc.  Only power-law fits that are based
on more than $nb=3$ log-mass bins are plotted.  The horizontal and
vertical lines, the faint open circles and open triangles have the
same meaning as in Fig.~\ref{fig:a_m_sv}.  }}
\label{fig:a_m_allst}
\vfill
\rotatebox{-90}{\resizebox{0.6 \textwidth}{!}
{\includegraphics{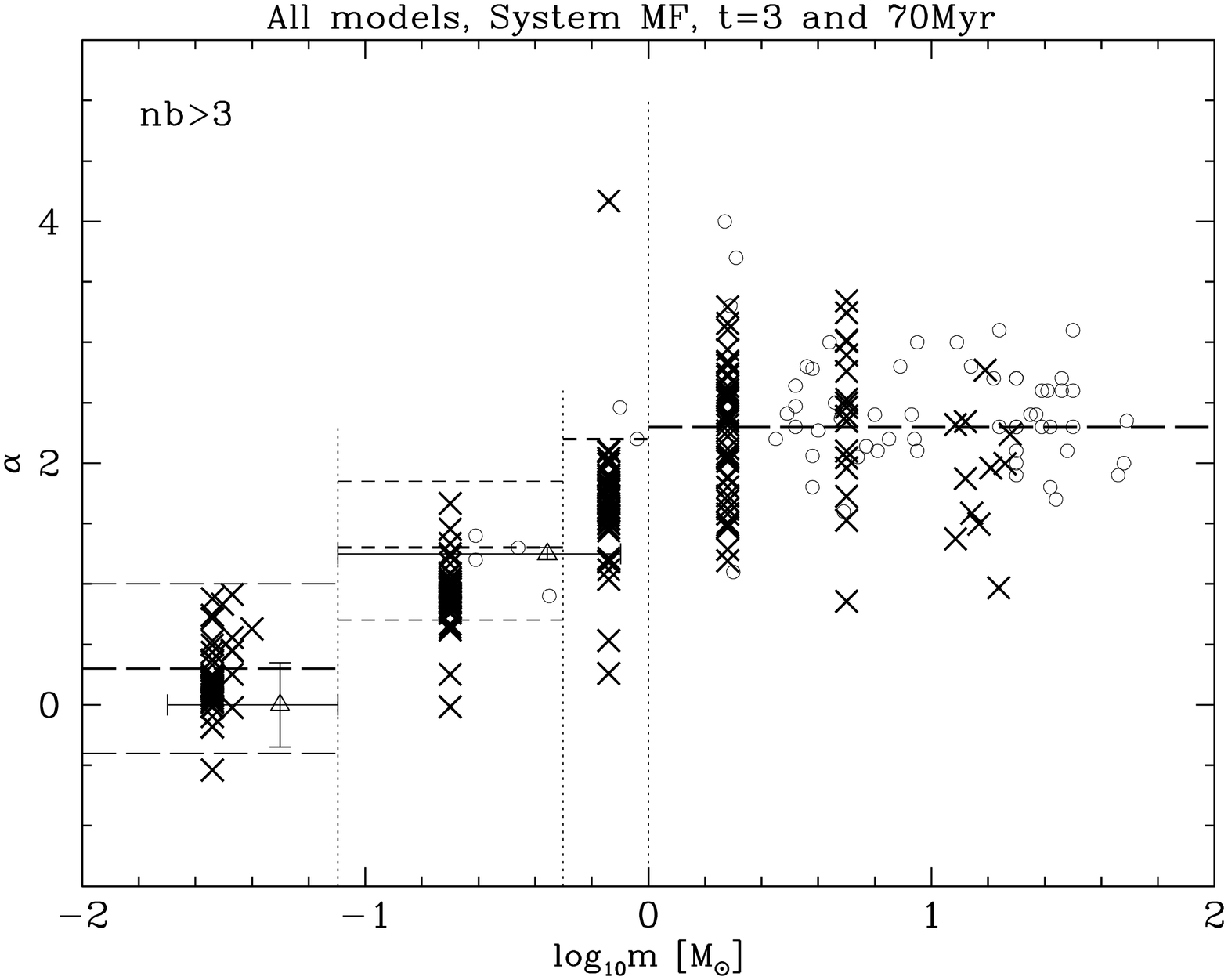}}}
\vskip 0mm
\caption
{\small{ As Fig.~\ref{fig:a_m_allst}, but assuming the observer cannot
resolve {\it systems}.}}
\label{fig:a_m_allsys}
\end{center}
\end{figure}

The model $\alpha$ values obtained by fitting power-laws to the model
system MFs are consistent with $\alpha\approx2.3$ for $m
\simgreat1\,M_\odot$, thus re-deriving the input IMF despite
unresolved binary systems. This result will be re-visited in future
work for the reasons stressed in Sections~\ref{sec:ms}}
and~\ref{sec:nb}.

For $m <1\,M_\odot$, the average system-$\alpha$ are too small,
except in the BD regime, where approximately the input value
($\alpha_0=+0.3$) is arrived at because of the small surviving binary
proportion. Fig.~\ref{fig:a_m_allsys} thus demonstrates that the
observational data (open circles and triangles) underestimate the
single-star power-law index in mass-ranges $b2$ and $b3$
(eq.~\ref{eq:bin}) by about $\Delta\alpha\approx0.5$, because binary
systems are not resolved. This is a {\it reliable result}, because of
the reasoning in Section~\ref{sec:nb}, that is, because the cluster
library used here has an extreme initial density. {\it Any Galactic
embedded cluster with a lower density may lead to a larger bias},
because in lower-density clusters the binary population is eroded at a
slower rate, allowing a higher binary proportion to survive for longer
times.  The binary proportion is certainly not lower in such clusters,
which is also confirmed by detailed analysis of observations
(e.g. K\"ahler 1999 for the Pleiades; Kroupa \& Tout 1992 for the
Praesepe).

Again, it is stressed that the above {\it corrections to $\alpha$ are
minimum values}, especially for BDs. The binary proportion of these
may be larger in clusters with lower density, because it takes longer
for $f_{\rm BD}$ to decrease in lower-density clusters.  The maximum
corrections to be applied to the observed, i.e. system MFs, are
derived from the models at $t=0$ (e.g. Fig.~\ref{fig:a_m_N1E4}):
$\Delta\alpha=+1.3$ for BDs, and $\Delta\alpha=+0.8$ for $0.08\le m <
1\,M_\odot$. Such large corrections are, however, unlikely, because
$f<1$ usually (except in Taurus--Auriga, cf. Luhman 2000). 

The observational data in Fig.~\ref{fig:a_m} therefore imply a
single-star IMF that is steeper than eq.~\ref{eq:imf} for
$0.08\simless m\simless 1\,M_\odot$ by $\Delta\alpha\approx0.5$ at
least.  Thus, for these data the corrected IMF has
$\alpha_1\approx1.8$ for $0.08-0.5\,M_\odot$, and $\alpha_2\approx2.7$
for $0.5-1\,M_\odot$, probably with unchanged $\alpha_0$ and
$\alpha_3$. The implications of this are discussed in
Sections~\ref{sec:revimf} and~\ref{sec:varimf}.

Figs~\ref{fig:a_m_allst} and~\ref{fig:a_m_allsys} show that the model
scatter in $\alpha$ is similar to that seen in the observational
sample.  Despite starting in each case with the same IMF, an observer
deduces power-law indices that have a scatter of about
$\sigma_\alpha=0.5$ for $m\simless1\,M_\odot$ and $\sigma_\alpha=1$
for $m\simgreat1\,M_\odot$, even if each stellar mass is measured
exactly. The finding is thus that the IMF can never be determined more
accurately than this scatter, and that the scatter seen in the
alpha-plot (Fig.~\ref{fig:a_m}) can be explained with Poisson noise
and stellar dynamical effects.

\section{DISCUSSION}
\label{sec:disc}
\noindent 
A cautionary remark concerning the alpha-plot is made, namely that in
reality the left and right parts of it are disjoint. Also, some
tentative evidence {\it for} a systematically varying IMF is
presented, especially in view of the proposed revised IMF.

\subsection{The dichotomy problem}
\label{sec:dich}
\noindent 
When considering the alpha-plot (Fig.~\ref{fig:a_m}), it must be
remembered that that left ($m\simless1\,M_\odot$) and right
($m\simgreat1\,M_\odot$) parts of it are actually disjoint.

That is, any nearby cluster that is older than a few~Myr so as to
allow the application of reasonably well understood pre-main sequence
or main sequence stellar models, contains no O~stars or is already too
old for them to still exist. This is very true for the Galactic-field
IMF -- there is only an indirect handle on $m\simgreat1\,M_\odot$
stars through stellar remnants, but this requires an excellent
understanding of stellar evolution, the sfh and Galactic-disk
structure (e.g. Scalo 1986).  Conversely, any population of stars for
which the MF is constrained through observations for $m>1\,M_\odot$ is
usually so far away that the left part of the alpha-plot is not
accessible to the observer, and/or so young that measuring the
derivative ($\alpha$), i.e. the {\it shape}, of the IMF for
$m<1\,M_\odot$ becomes a lottery game because of the uncertain
pre-main sequence tracks (Section~\ref{sec:intro}).

That low-mass stars do form in large numbers in any population that
also forms O~stars {\it is} established. Examples are the ONC
(Hillenbrand 1997), R136 in the 30~Dor region in the LMC (Siriani et
al. 2000), and NGC~3603, the most massive visible HII region in the MW
(Brandl et al. 1999). However, the ONC is so young that mass estimates
become unreliable, compromising conclusions about the detailed shape
of its IMF, and in the other cases the census of low-mass stars is not
complete.  Thus, the shape of the IMF spanning log$_{10}m=-2$ to~2 is
not known for {\it any} population, and it remains an act of faith to
assume that the IMF can be approximated by the form of
eq.~\ref{eq:imf}.

Globular clusters consist entirely of low-mass stars today, but the
existence of neutron stars demonstrates that massive stars formed in
them as well.  Paresce \& De Marchi (2000) suggest that the MF for a
sample of a dozen globular clusters can be fit by a log-normal MF with
approximately one characteristic stellar mass and standard
deviation. A further analysis will show how the differences compare
with the spread in $\alpha$ seen in Fig.~\ref{fig:a_m}.  More
interesting in the present context is that Piotto \& Zoccali (1999),
who use the same stellar models by Baraffe et al. (1997) as Paresce \&
De Marchi, demonstrate that power-law MFs fit rather well for a wide
range of globular clusters, with $\alpha\approx0.5-1.2$ for
$m\simless0.5-0.7\,M_\odot$, but the IMF is not measurable for
stars with $m\simgreat 0.7\,M_\odot$.

\subsection{A revised IMF}
\label{sec:revimf}
\noindent 
In Section~\ref{sec:amsyn} the suggestion is made that the systematic
bias towards low $\alpha_1$ and $\alpha_2$ due to unresolved binaries
implies that the single-star IMF may be steeper than inferred from
observations that do not resolve binary systems.  Correcting the
ensemble of observed $\alpha$ in Fig.~\ref{fig:a_m} for this bias
leads to the following revised IMF,
\begin{equation}
          \begin{array}{l@{\quad\quad,\quad}l}
\alpha_0 = +0.3\pm0.7   &0.01 \le m/M_\odot < 0.08, \\
\alpha_1 = +1.8\pm0.5   &0.08 \le m/M_\odot < 0.50, \\
\alpha_2 = +2.7\pm0.3   &0.50 \le m/M_\odot < 1.00, \\
\alpha_3 = +2.3\pm0.7   &1.00 \le m/M_\odot,\\
          \end{array}
\label{eq:imfr}
\end{equation}
where the uncertainties from eq.~\ref{eq:imf} are carried over. 

The revised IMF has, for stars with $0.01\le m\le50\,M_\odot$, an
average stellar mass $<m>=0.20\,M_\odot$ and leads to the following
population: 50~\% BDs ($0.01-0.08\,M_\odot$) contributing 10~\% to the
stellar mass, 44~\% M~dwarfs ($0.08-0.5\,M_\odot$) contributing 39~\%
mass, 4.3~\% ``K''~dwarfs ($0.5-1.0\,M_\odot$) contributing 14~\%
mass, 2.3~\% ``intermediate mass (IM) stars'' ($1.0-8.0\,M_\odot$)
contributing 24~\% mass, and 0.15~\% ``O'' stars ($>8\,M_\odot$)
contributing 12~\% mass. O and IM stars thus contribute together
36~per cent of the total mass.  If $\alpha_4=1.15$ ($m>8\,M_\odot$)
then 50~per~cent of the mass is in stars with $8\le m\le120\,M_\odot$.

This revised IMF can be viewed as the {\it present-day star-formation
IMF}, and is in good agreement with the pre-stellar clump MF measured
by Motte et al. (1998) and Johnstone et al. (2000) for $\rho$~Oph:
$\alpha_1\approx1.5$ and $\alpha_2\approx2.5$; especially so since
each clump is likely to form a multiple star.

\subsection{Possible evidence for a variable IMF}
\label{sec:varimf}
\noindent 
A short account is made of the most promising evidence for a
systematically varying IMF. The discussion in Sections~\ref{sec:gl}
to~\ref{sec:wd} is visualised in Fig.~\ref{fig:imfv}, in which the
various IMFs are compared.

\begin{figure}
\begin{center}
\rotatebox{0}{\resizebox{0.77 \textwidth}{!}
{\includegraphics{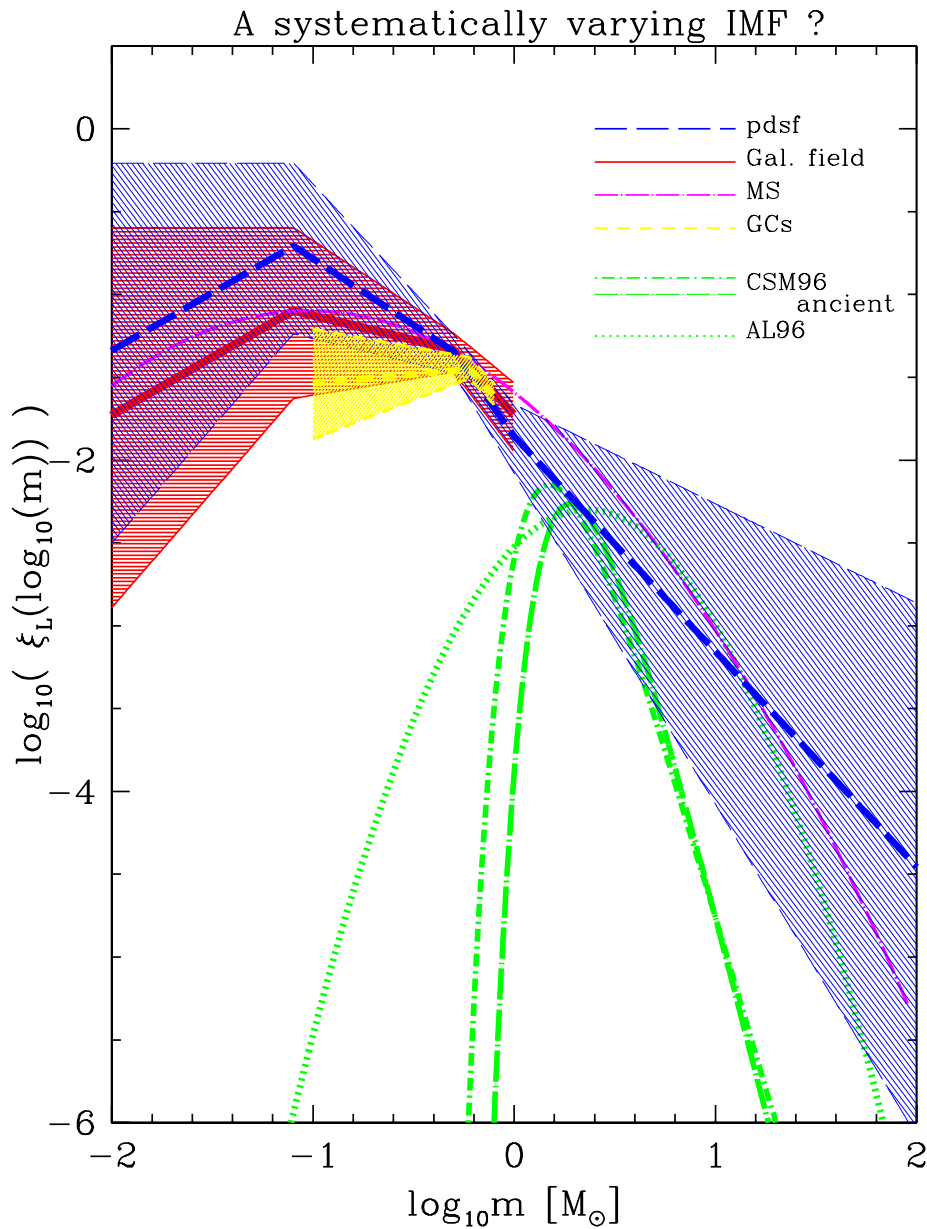}}}
\vskip 0mm
\caption
{\small{ Evidence for a systematically changing IMF. The present-day
star-formation (pdsf) IMF (eq.~\ref{eq:imfr}) is shown as the thick
dashed line.  The Galactic-field IMF (eq.~\ref{eq:imf}) is the thick
solid line. It is truncated at $m=1\,M_\odot$ to express our ignorance
about the IMF for $m>1\,M_\odot$ for this population that has an
average age of about~5~Gyr (the ``dichotomy problem'',
Section~\ref{sec:dich}).  In both cases the shaded areas represent the
approximate 95--99~per cent confidence region. For comparison, the
Miller \& Scalo (1979) log-normal IMF for a constant star-formation
rate and a Galactic disk age of 12~Gyr is plotted as the thin
long-dash-dotted curve (it's derivative is shown in
Fig.~\ref{fig:a_m}). Seven globular clusters give $\alpha_1=0.89$,
with upper and lower values of~1.22 and~0.53, and $\alpha_2=2.3$ for
$0.6<m<0.8\,M_\odot$ (Piotto \& Zoccali 1999) as indicated by the
short-dashed lines and the heavily shaded area. Three possible IMFs
for Galactic-halo WD-progenitors are suggested by the thick
long-dash-dotted and short-dash-dotted lines (Chabrier, Segretain \&
M\'era 1996, CSM96), and the thick dotted line (Adams \& Laughlin
1996, AL96).  The MFs have been scaled such that they agree near
$0.5\,M_\odot$, except for the ancient IMFs, which are scaled
to fit the Galactic-field IMF near $2\,M_\odot$.  }}
\label{fig:imfv}
\end{center}
\end{figure}

\subsubsection{Globular clusters vs Galactic field}
\label{sec:gl}
\noindent 
The suggestion in Section~\ref{sec:revimf} that the alpha-plot
(Fig.~\ref{fig:a_m}) may imply a present-day star-formation (pdsf) IMF
(eq.~\ref{eq:imfr}) that is steeper than the Galactic-field IMF
(eq.~\ref{eq:imf}) is interesting when compared to the MFs estimated
for globular clusters (Section~\ref{sec:dich}).  These are very
ancient and metal poor systems, so that a systematically different IMF
(Larson 1998) ought to be manifest in the data. The difference should
be in the sense that globular clusters ought to contain a
characteristic stellar mass that is larger than that in more
metal-rich populations.  The systematically flatter MF in globular
clusters compared to the Galactic-field IMF (eq.~\ref{eq:imf}), and
especially to the pdsf IMF (eq.~\ref{eq:imfr}), may thus be due to a
real difference in the star-formation conditions. 

However, unfortunately the evidence is not conclusive because globular
clusters have lost preferentially low-mass stars, leading to a
systematic flattening of the MF with time, unless the clusters are at
large Galactocentric distances (Vesperini \& Heggie 1997). The binary
proportion in globular clusters is typically smaller ($f\simless0.3$)
than in the Galactic field ($f\approx0.6$) but probably not negligible
(Hut et al. 1992; Meylan \& Heggie 1997), and correction for their
effects may also steepen the measured MF. Approximate corrections that
increase the measured $\alpha$ are $\Delta\alpha<1$ for dynamical
evolution (fig.~6 in Vesperini \& Heggie 1997) and
$\Delta\alpha\approx0.2$ for unresolved binary systems, but a
case-by-case study is required for detailed estimates. In their
sample, Piotto \& Zoccali (1999) find evidence for flatter MFs at
smaller Galactocentric distances suggesting loss of low-mass stars as
being an important bias. But, there is also evidence for a correlation
such that more metal-rich clusters have larger $\alpha$.

The Galactic-field IMF (eq.~\ref{eq:imf}) is valid for stars that are,
on average, about 5~Gyr old, and which were formed at a different
epoch of Galactic evolution than the stars in the clusters featuring
in Fig.~\ref{fig:a_m}.  This, then, suggests a possible systematic
shift of star formation towards producing relatively more low-mass
stars as star-formation moves towards conditions that may favour lower
fragmentation masses through higher metallicities and lower cloud
temperatures. That the pre-stellar core MF in $\rho$~Oph is somewhat
steeper than the Galactic-field IMF (eq.~\ref{eq:imf}), while being
consistent with a fragmentation origin (Motte et al. 1998), supports
this notion.

\subsubsection{Galactic-halo white dwarfs}
\label{sec:wd}
\noindent 
Another possible empirical hint for a variable IMF may be provided if
part of the dark halo of the Galaxy were in the form of ancient white
dwarfs.  This is becoming a distinct possibility, given that a handful
of candidate ancient halo white dwarfs have been discovered (Elson,
Santiago \& Gilmore 1996; Ibata et al. 1999, 2000; M\'endez \& Minniti
2000).

From eq.~\ref{eq:imf} one obtains per WD progenitor ($1-8\,M_\odot$,
e.g. Weidemann 1990) about~8 dwarfs with $m=0.1-0.7\,M_\odot$. No such
halo dwarfs that might belong to the same population as the putative
WDs have been found, requiring a radically different IMF for their
progenitor stars than is seen today in the Galactic disk. Also, for
consistency with chemical enrichment data, such an IMF cannot have
many stars with $m\simgreat5\,M_\odot$ (Chabrier, Segretain \& M\'era
1996; Adams \& Laughlin 1996; Larson 1998; Chabrier 1999).

\subsubsection{Radial variation in a very young cluster}
\label{sec:radvar}
\noindent 
Hillenbrand (1997) demonstrates that the ONC has pronounced mass
segregation, and this may be interpreted as an IMF which has a radial
variation, {\it if} dynamical mass segregation is not fast enough to
produce such mass segregation within the age of the cluster. The age
of the ONC is estimated to be less than 1~Myr for most ONC stars
(Hillenbrand 1997; Palla \& Stahler 1999), and Bonnell \& Davies
(1998) suggest, by using a softened $N$-body code, that mass
segregation takes too long to produce the observed effect. However,
stellar-dynamical computations with a direct $N$-body code that
correctly treats the many close encounters must be applied to this
problem (Kroupa, in preparation). If the mass-segregation time-scale
is too long to produce the observed effect, then we would have a
well-documented case of a variable IMF most likely through
interactions of pre-stellar cores, as suggested by Bonnell et
al. (1998) and Klessen (2001).

\section{CONCLUSIONS}
\label{sec:concl}
\noindent
The following three main points are covered in this paper: 

\noindent{\bf I.} \underbar{The Galactic-field IMF.}  The form of the
average IMF consistent with constraints from local star-count data and
Scalo's (1998) compilation of MF power-law indices for young clusters
and OB associations is inferred.  The IMF is given by
eq.~\ref{eq:imf}. This form may be taken as the universally valid IMF.

\noindent{\bf II.} \underbar{The alpha-plot: scatter and systematics.}
Assuming the universal IMF (eq.~\ref{eq:imf}), how large are the apparent
variations produced by Poisson noise, the dynamical evolution of
star-clusters and unresolved binary systems?

This is studied by making use of the alpha-plot, in which IMF
power-law indices inferred for $N$-body model populations are plotted
as a function of stellar mass.  The extreme assumption is made that
the observer can measure each stellar or binary-system mass
exactly. The resultant {\it apparent variation} of the IMF thus
defines the {\it fundamental limit for detecting true variations}. Any
true variation of the IMF that is smaller than this fundamental limit
cannot be detected. This is the reason why no robust evidence for a
variable IMF has surfaced to date. The available population samples
are too small (e.g. {\it one} ONC vs {\it one} $\rho$~Oph).

The model clusters have an initial binary proportion of unity and
contain $N=800, 3000$ and~$10^4$ stars with a central density as in
the ONC.  Clusters with a smaller initial density evolve on a longer
time-scale. The binary-star problem is thus potentially worse in
less-dens clusters, because binary systems survive for longer.

The observed spread of power-law indices is arrived at
approximately. For the ensemble of model clusters studied here it is
$\sigma_\alpha\approx0.7$ for BDs, $\sigma_\alpha\approx0.5$ for stars
in the mass range $0.1-1\,M_\odot$, and $\sigma_\alpha\approx1$ for
stars with $m\simgreat1\,M_\odot$ (Fig.~\ref{fig:a_m_allsys}).  

For stars with $m\simgreat1\,M_\odot$, the system MF has, on average,
the same power-law index as the underlying single-star IMF.  That is,
the present models do not lead to any systematic bias in this mass
range (but see caveat in Section~\ref{sec:ms}).  Similarly, for BDs
the input $\alpha$ is arrived at in the mean, but only if the
population is at least a few crossing times old, because by then most
BD binaries and star--BD binaries have been disrupted.  For a
dynamically younger population, $\alpha$ (and the number of BDs) will
be underestimated depending on the binary proportion.

To correct for unresolved binaries, the measured power-law index has
to be increased by $0\simless \Delta\alpha_0 \simless 1.3$ for BDs and
$0.5\simless \Delta\alpha_{1,2} \simless 0.8$ for $0.08\le
m\le1\,M_\odot$, the upper and lower limits applying for clusters that
are unevolved ($t=0$) and a few crossing times old, respectively,
assuming $f=1$ when $t=0$.  For a population in a cluster that is a
few crossing times old, the corrections reduce to
$\Delta\alpha_0\approx0$ and $\Delta\alpha_{1,2}\approx0.5$.  These
corrections have to be applied to any young population to infer the
single-star IMF.

Finally, as a cautionary remark, the left and right parts of the
alpha-plot are observationally disjoint.  It is an act of faith to
assume that $\alpha(m)$ has the smooth dependence given by
eq.~\ref{eq:imf}.  

\noindent{\bf III.} \underbar{IMF variations.}  Applying the above
corrections to the ensemble of observed young clusters, a revised (or
present-day star-formation) IMF is arrived at (eq.~\ref{eq:imfr}). It
is steeper for $m\simless 1\,M_\odot$ than the Galactic-field IMF
(eq.~\ref{eq:imf}), which is a mixture of star-formation events with
an average age of about 5~Gyr. The pre-stellar clump mass-spectrum in
the present-day star-forming cloud $\rho$~Oph (Motte et al. 1998;
Johnstone et al. 2000) also indicates a steeper single-star MF than
the Galactic-field MF.  Intriguingly, the ancient MFs in globular
clusters have $\alpha\simgreat0$, but closer to~0 than the
Galactic-field IMF. The recent detection of candidate white dwarfs in
the Galactic halo suggests that the IMF of the progenitor population
must have been radically different by producing few if any low-mass
and massive stars ($\alpha<<0$ for $m\simless0.5\,M_\odot$ and
$\alpha>>0$ for $m\simgreat2\,M_\odot$).

Furthermore, the well-developed mass segregation in the very young
($\simless 2$~Myr) ONC may exemplify a locally radially-varying IMF,
{\it if} dynamical mass segregation is too slow. If $N$-body
calculations confirm this to be the case (work is in progress), then
the ONC will be definite proof that the local conditions determine the
average stellar mass, rather than it merely being the result of
statistical fluctuations.

\vfill

The tentative suggestion is thus that some systematic variation may
have been detected, with star-formation possibly producing relatively
more low-mass stars at later Galactic epochs. Such a variation would
be expected in the mass range ($\simless1\,M_\odot$) in which
turbulent fragmentation, which depends on the cooling rate and thus
metal abundance, dominates. Future observations of LMC populations
might verify if the IMF has systematically smaller $\alpha$ for
$m\simless 1\,M_\odot$ than the Galactic-field or present-day
star-formation IMF.  Unfortunately though, even if there is a trend
with metallicity, it will be very arduous to uncover a systematic
difference in $\alpha$ between the MW and LMC at low masses because
the metallicity difference is not very large while the
$\alpha$-scatter is.  A lack of systematic differences in $\alpha$ for
$m\simgreat10\,M_\odot$ between MW and LMC populations may be a result
of one physical mechanism, such as coalescence, dominating in the
assembly of massive stars (Larson 1999).

\newpage
\acknowledgements 
\vskip 10mm
\noindent{\bf Acknowledgements}
\vskip 3mm
\noindent 
I am grateful to John Scalo for letting me have his compilation of
mass function power-law indices, and I thank Sverre Aarseth for making
{\sc Nbody6} freely available. The calculations were performed at the
Institute for Theoretical Astrophysics, Heidelberg University, where I
spent a few very pleasant years.  I acknowledge support through DFG
grant KR1635.

%

\end{document}